\documentclass{aa}  
\usepackage{color}
\usepackage{graphicx}
\usepackage{multirow}
\usepackage{txfonts}

\begin{document} 

        \title{2.5D global-disk oscillation models of the Be shell star $\zeta$\,Tauri} 
                \subtitle{I. Spectroscopic and polarimetric analysis}

\author{    C. Escolano \inst{1}
        \and A. C. Carciofi \inst{1}
        \and A. T. Okazaki \inst{2}
        \and T. Rivinius \inst{3}
        \and D. Baade \inst{4}
        \and S. {\v S}tefl \inst{5}\thanks{deceased}}

\institute{ Instituto de Astronomia, Geof\'isica e Ciencias Atmosf\'ericas,
        Universidade de S\~ao Paulo, Rua do Mat\~ao 1226,
        Cidade Universit\'aria, S\~ao Paulo, SP 05508-090, Brazil \\
\email{cyril.escolano@usp.br}
        \and Hokkai-Gakuen University, Toyohira-ku, Sapporo 062-8605, Japan
        \and ESO -- European Organisation for Astronomical Research in the Southern Hemisphere,
        Casilla 19001, Santiago, Chile
        \and ESO -- European Organisation for Astronomical Research in the Southern Hemisphere,
        Karl-Schwarzschild-Str. 2, Garching, Germany
        \and ALMA -- Atacama Large Millimeter Array, Alonso de C{\' o}rdova 3107, Vitacura, Santiago, Chile}

\date{Received 02 December 2014 / Accepted 04 February 2015}

\abstract
{A large number of Be stars exhibit intensity variations of
their violet and red emission peaks in their \ion{H}{i} lines observed in
emission. This is the so-called $V/R$ phenomenon, usually explained by the
precession of a one-armed spiral density perturbation in the circumstellar disk. That global-disk oscillation
scenario was confirmed, both observationally and theoretically, in the previous series of two papers analyzing
the Be shell star $\zeta$\,Tauri. The vertically averaged (2D) global-disk oscillation model used at the time
was able to reproduce the $V/R$ variations observed in H$\alpha$, as well as the spatially resolved interferometric
data from AMBER/VLTI. Unfortunately, that model failed to reproduce the $V/R$ phase of \ion{Br}{15} and
the amplitude of the polarization variation, suggesting that the inner disk structure predicted by the model
was incorrect.}
{The first aim of the present paper is to quantify the temporal variations of the shell-line characteristics of
$\zeta$\,Tauri. The second aim is to better understand the physics underlying the $V/R$ phenomenon by
modeling the shell-line variations together with the $V/R$ and polarimetric variations. The third aim is to test a
new 2.5D disk oscillation model, which solves the set of equations that describe the 3D perturbed disk
structure but keeps only the equatorial (\emph{\emph{i.e.}}, 2D) component of the solution. This approximation
was adopted to allow comparisons with the previous 2D model, and as a first step toward a future 3D
model.}
{We carried out an extensive analysis of $\zeta$\,Tauri's spectroscopic variations by measuring various quantities
characterizing its Balmer line profiles: red and violet emission peak intensities (for H$\alpha$, H$\beta$, and \ion{Br}{15}),
depth and asymmetry of the shell absorption (for H$\beta$, H$\gamma$, and H$\delta$), and the respective position
(\emph{\emph{i.e.}}, radial velocity) of each component. We attempted to model the observed variations by implementing in
the radiative transfer code HDUST the perturbed disk structure computed with a recently developed 2.5D global-disk
oscillation model.}
{The observational analysis indicates that the peak separation and the position of the shell absorption both exhibit
variations following the $V/R$ variations and, thus, may provide good diagnostic tools of the global-disk oscillation phenomenon.
The shell absorption seems to become slightly shallower close to the $V/R$ maximum, but the scarcity of the data does not allow
the exact pattern to be identified. The asymmetry of the shell absorption does not seem to correlate with the $V/R$ cycle; no
significant variations of this parameter are observed, except during certain periods where H$\alpha$ and H$\beta$ exhibit
perturbed emission profiles. The origin of these so-called triple-peak phases remains unknown. On the theoretical side, the
new 2.5D formalism appears to improve the agreement with the observed $V/R$ variations of H$\alpha$ and \ion{Br}{15},
under the proviso that a  large value of the viscosity parameter, $\alpha = 0.8$, be adopted. It remains challenging for
the models to reproduce consistently the amplitude and the average level of the polarization data. The 2D formalism provides
a better match to the peak separation, although the variation amplitude predicted by both the 2D and 2.5D models is smaller than the
observed value. Shell-line variations are difficult for the models to reproduce, whatever formalism is adopted.}
{}


\maketitle

\section{Introduction}

Following \citeauthor{struve31}'s (\citeyear{struve31}) picture, shell stars are classical Be stars
observed with their disks edge-on.  The circumstellar disks lie between the observer
and the central star (whereas the disks do not project on the photosphere in the case of 
\emph{\emph{classical}} Be stars) and this results in distinct phenomenology. For instance, their spectra exhibit
sharp absorption profiles in Balmer lines and in lines from heavier elements that are ionized once. These
so-called shell lines are expected to be formed in the innermost section of the rotating disk. Their
narrow profiles are the consequence of the extremely small differential velocity projected in the
line of sight toward the star (where the rotation is perpendicular to the line of sight).

In the simplest case of an axisymmetric, unperturbed disk, both emission and shell-line profiles
would be time-invariant. This is certainly the case for some well-known objects like $\beta$\,CMi
for the classical Be stars or $o$\,Aqr for the shell stars \citep{hanuschik96a, rivi06}. However, some
shell stars exhibit large variations in the shape, intensity, and radial velocity of their central core,
manifesting the existence of time-dependent asymmetries and non-azimuthal velocity fields in the
disk. As such, shell lines are excellent diagnostic tools to probe Be disk dynamics, as was proven
by the analyses performed by \citet{stefl12} on 48\,Lib and by \citet{mon13} on EW\,Lac.

The shell star $\zeta$\,Tauri (HR\,1910, HD\,37202) has been widely studied in the past. The first
reason is probably because it is a bright object, whose ideal position in the sky makes it observable
from both the northern and southern hemispheres, thus providing a rich dataset. The second and certainly
most important reason is because $\zeta$\,Tauri exhibits various peculiarities that make it the perfect
laboratory for understanding the physics of Be stars and for testing theoretical models. Its most important
feature is the so-called $V/R$ variation, \emph{\emph{i.e.}}, the often cyclic fluctuation  of the ratio between
the violet and red emission peak intensities, in \ion{H}{i} lines observed in emission. These variations are
thought to be the manifestation of a global-disk oscillation \citep{kato83}, provoking important modifications
to the density and velocity structures in the disk.

In \citet[hereafter ZT\,I]{stefl09}, a large pool of spectroscopic and spectro-interferometric data covering
three complete $V/R$ cycles was gathered and analyzed. In \citet[ZT\,II]{carciofi09}, a vertically averaged
(hereafter 2D) global-disk oscillation model of \citet{okazaki97} and \citet{papaloizou92} was combined
with the radiative transfer code HDUST to model this dataset. They successfully reproduced the observed
$V/R$ cycle of H$\alpha$, as well as the spatially resolved AMBER/VLTI interferometric data, providing the
first observational and theoretical confirmation of the existence of global density waves. However, the models
shown in ZT\,II were not able to reproduce the $V/R$ cycle of \ion{Br}{15} and its polarimetric variations. These
two observables originate in the inner disk; it was thus hypothesized at the time that the model geometry in these
regions was incorrect. A possible solution emerged with the 3D disk oscillation model of \citet{ogilvie08}, who
has shown that accounting for the vertical structure of the perturbation strongly modifies  the properties of the
solution for the perturbation pattern in the disk.

There are multiple objectives of the present paper. The first  is to quantify the shell-line variations of
$\zeta$\,Tauri that have been, until now, only considered in first order, global quantities. We thus propose
a  finer grained quantitative description of various aspects of the line profile, not limited to the emission
components alone. By doing so, we can assess the diagnostic potential of various line features for probing the
disk structure and dynamics. The second objective is to model these variations and test whether or not the
$V/R$ parameters determined in ZT\,II are consistent with the observational diagnostics introduced here.
The final objective is to test the model of \citet{ogilvie08} in a modified \emph{\emph{2.5D}} version and compare its
predictions with the former 2D formalism.

In Sect.~2, we briefly present the observational data and give some details about the various line characteristics
we measured. Section~3 is dedicated to the quantitative description of the observed temporal shell-line variations
and their relation with the $V/R$ cycles. In Sect.~4, global-disk oscillation models will be compared to observations.
Before concluding in Sect.~6, the structure of the models and the effect of some model parameters on the
observables will be discussed in Sect.~5.

\begin{figure}[ht]
        \centering
        \includegraphics[trim=0.cm 0.cm 0.cm 0.cm, clip=true, totalheight=0.20\textheight]{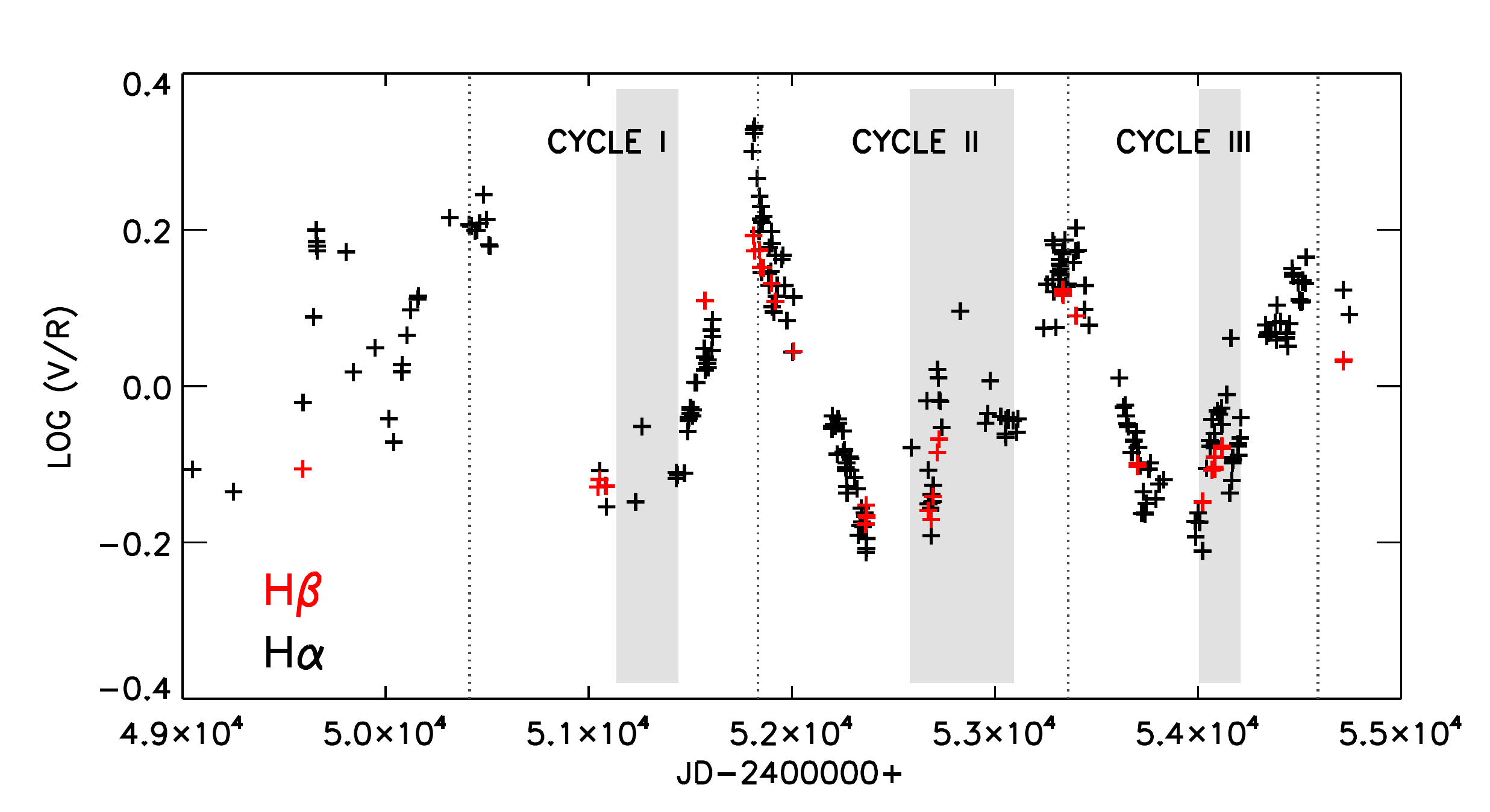} \\
        \includegraphics[trim=0.cm 0.cm 0.cm 0.cm, clip=true, angle=90., totalheight=0.20\textheight]{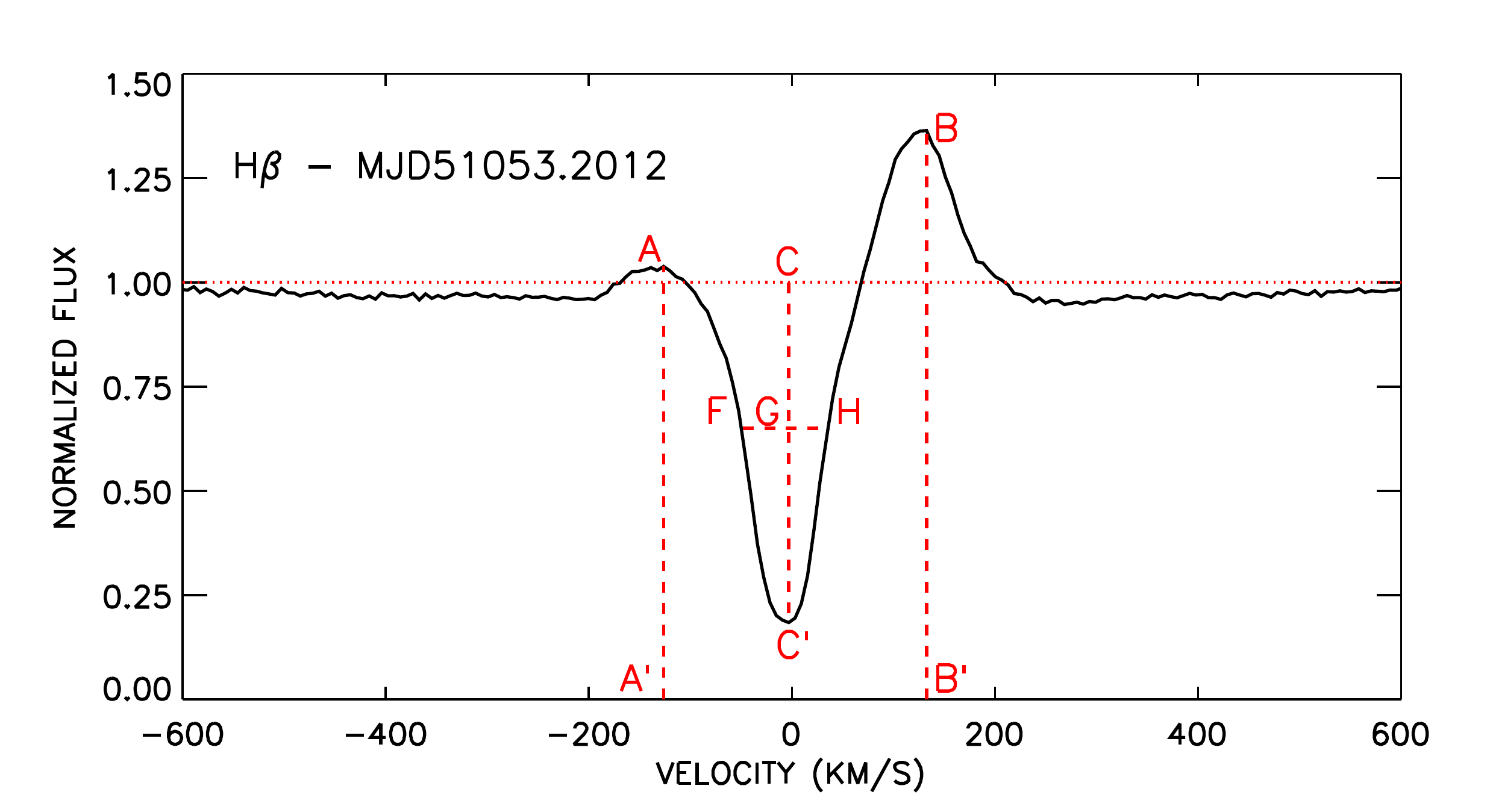}
        \caption{
        \emph{Top:} H$\alpha$ (black) and H$\beta$ (red) $V/R$ curves. Three complete $V/R$ cycles were
        covered between 1997 and 2010. The gray regions  indicate the triple-peak phase for each cycle.
        \emph{Bottom:} Measured quantities illustrated on a H$\beta$ line profile: violet ($AA'$) and red ($BB'$)
        emission peaks, depth of the central absorption ($CC'$) and asymmetry ($FG/FH$ ratio). The respective
        radial velocities of the three components (emission peaks and central absorption) were measured
        whenever possible. The horizontal line marks the normalized continuum.}
\label{fig1}
\end{figure}

\begin{figure*}[ht]
        \centering
        \includegraphics[trim=0.cm 0.5cm 0.cm 0.cm, clip=true, angle=0, scale=0.7]{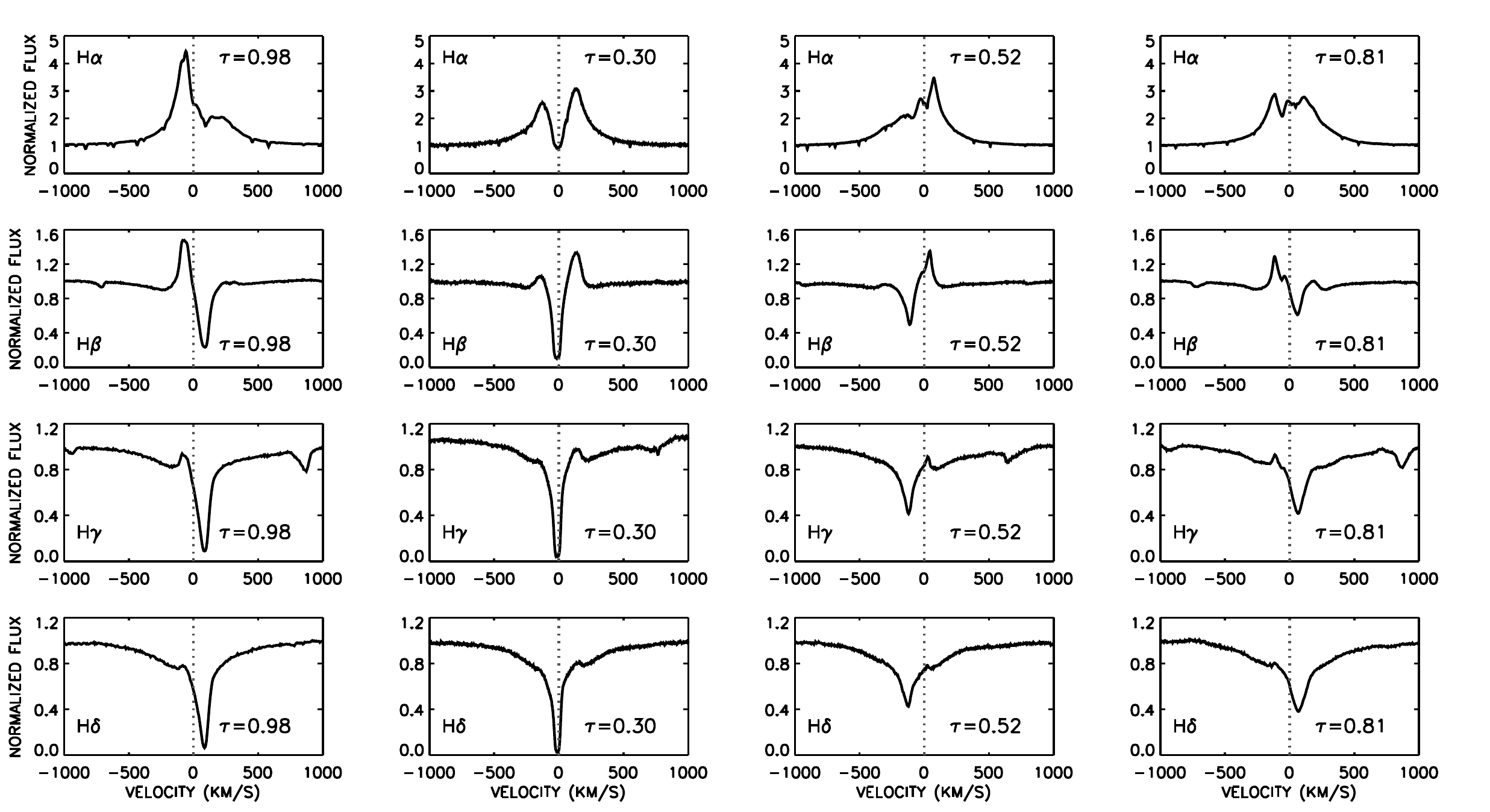}
        \caption{Optical spectrum of $\zeta$\,Tauri observed at different phases of the $V/R$ cycle.
        From left to right are shown the $V/R$ phases $\tau=$0.98, 0.30, 0.52, and
        0.81; from top to bottom are shown the successive Balmer lines from H$\alpha$ to H$\delta$.
        The vertical dotted line marks the rest velocity ($v_{rad} = 0$ km\,s$^{-1}$).}
\label{fig2}
\end{figure*}

\section{Observational data and analysis method}

We focus our analysis on the hydrogen lines across the optical and near-IR ranges because
these are the lines our numerical tools can model (see next section). The observational material
is the same as that used in ZT\,I and ZT\,II. It consists of around 300 spectra, combining
observations of our own as well as archival data from ESO and from the BeSS database
\citep{neiner07}. The dataset includes three complete $V/R$ cycles of approximately four years
each observed between 1997 and 2010. An inspection of the BeSS data observed after 2010 shows
that the strength of the disk emission has drastically decreased and at the same time the $V/R$ cycle
has ceased so that $V/R \approx 1$ since then.

Figure~1 (top panel) shows the $V/R$ curve for H$\alpha$ and H$\beta$ between 1997
and 2010, highlighting the quasi-cyclic nature of the phenomenon. The regions  in gray correspond
to the triple-peak phases (see below). Although the three cycles (labeled I, II, and III) have slightly different
lengths, we followed the definition of ZT\,I for the $V/R$ phase $\tau$, taking the origin at the very first $V/R$
maximum (MJD~50414) and an average cycle length of 1429 days.

Figure~2 presents a sample of Balmer lines observed at different $V/R$ phases; the vertical dotted
lines mark the rest wavelength (null radial velocity). The most striking features are the following:

\begin{itemize}
        \item $\tau=0.98$: This phase corresponds to the $V/R$ maximum. The violet emission peak
        is extremely strong in H$\alpha$ and H$\beta$ while it is much weaker, but still
        observable, in H$\gamma$ and H$\delta$. The red emission component is absent
        from the last two lines. The central shell absorptions are all redshifted.
        \newline
        
        \item $\tau=0.30$: These profiles are observed at a phase where $V/R \approx 1$, moving toward
        the $V/R$ minimum. Red emission peaks are a bit stronger than the violet ones in H$\alpha$ and
        H$\beta$. A  faint emission is observed on the violet side of H$\gamma$ and H$\delta$. The shell
        profiles are all centered on the rest velocity ($v_{rad}=0$ km\,s$^{-1}$) and are deeper than in any other
        phase.
        \newline

        \item $\tau=0.52$: This phase corresponds to the $V/R$ minimum; the observed features are qualitatively
        the opposite to the ones described for $\tau=0.98$: the red emission peaks are stronger than the violet ones
        and the shell absorptions are blueshifted with respect to the rest velocity. Additionally, the shell components
        appear broader and much shallower. We note  that all the lines exhibit a somewhat perturbed profile on their
        red side. For instance,  H$\alpha$  presents a central depression in its red emission component and, at nearly
        the same radial velocity, H$\beta$, H$\gamma$, and H$\delta$ present a peak. These specific features will
        be  discussed in Sect.~5.3.
        \newline

        \item $\tau=0.81$: This phase illustrates very well the difficulty of measuring the $V/R$ ratio and the peak
        radial velocities during the transition from $V < R$ to $V \approx R$ (triple-peak phase), as H$\alpha$
        exhibits three emission peaks of comparable intensity. Meanwhile, a small dip appears in the H$\beta$
        and H$\gamma$ blue emission peaks. The central shell absorptions are all redshifted and are shallower
        than during previous $V/R$ phases.
\end{itemize}

This quick overview provides a qualitative description of the phenomenon. We performed a more quantitative
approach by measuring  various line profile parameters for each spectrum, illustrated on an H$\beta$ profile
in Fig.~1 (bottom panel). The $V/R$ values plotted below for H$\alpha$ and \ion{Br}{15 } are those from ZT\,I, they
were completed by new measurements on H$\beta$, H$\gamma$, and H$\delta$. All the quantities
were derived by fitting a Gaussian curve to the top ($resp.$ bottom) of the violet and red emission peaks ($resp.$
central absorption). In the case of a triple-peaked  emission profile, the Gaussian was applied to the red component
as a whole without considering the additional absorption, making the measurements more uncertain in these
specific cases. 

\begin{itemize}

        \item \textbf{\emph{V/R} ratio:} The $V/R$ ratio can be defined  in two different ways, the first
         considering the peak height above the continuum level (normalized to unity in the present study),
        the second  considering the peak height above the zero flux level. That second  more
        straightforward definition was adopted here. Hence, the $V/R$ values correspond to the ratio between $AA'$
        ($V$) and $BB'$ ($R$).
        \newline

        \item \textbf{Depth of the central absorption:} The depth of the central core ($CC'$) was defined
        as the difference between the local continuum level (equal to unity) and the minimum value at the
        bottom of the line.
        \newline

        \item \textbf{Position/Radial velocity:} The respective heliocentric radial velocities of the central absorption
        ($V_{c}$ at $CC'$), violet ($V_{v}$ at $AA'$), and red ($V_{r}$ at $BB'$) emission peaks were measured.
        \newline

        \item \textbf{Asymmetry:} Based on the recent work of \citet{mon13}, we introduced a parameter to quantify
        the degree of asymmetry of the shell absorption. This parameter is defined as the ratio between the width of
        the blue wing at half maximum ($FG$) measured from the line center and the FWHM of the central absorption
        ($FH$). A value of 0.5 corresponds to a symmetric profile. Higher values indicate that the blue edge is broader
        (\emph{\emph{i.e.}}, farther from the line center) than the red edge; smaller values indicate that the red edge is the
        broader one.
        \newline

\end{itemize}

\newpage
\begin{figure}[!h]
\centering
        \includegraphics[trim=0cm 0cm 0.5cm 0.5cm, clip=true, angle=90, width=0.41\textwidth]{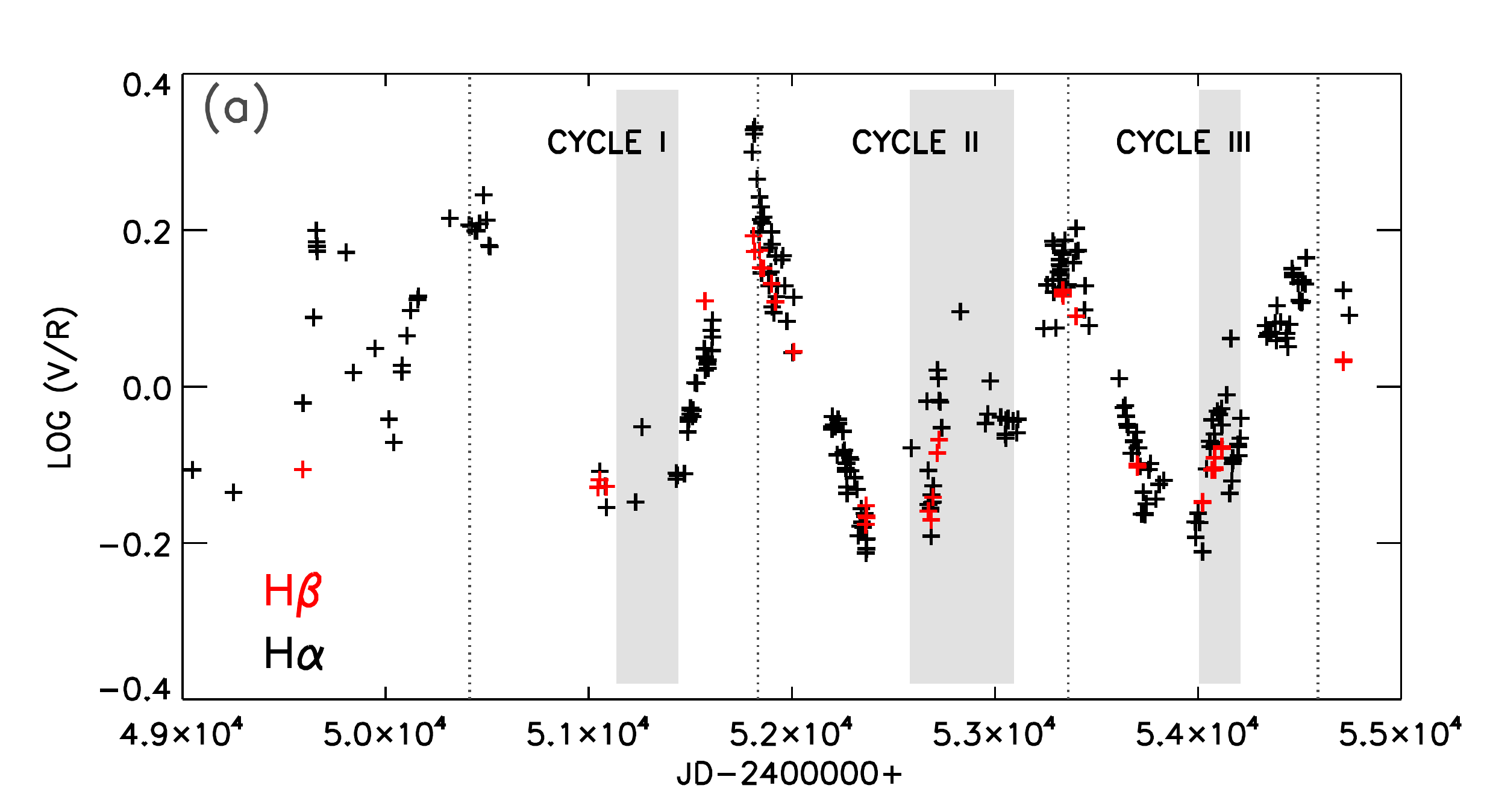} \\
        \includegraphics[trim=0cm 0cm 0.5cm 0.5cm, clip=true, angle=90, width=0.41\textwidth]{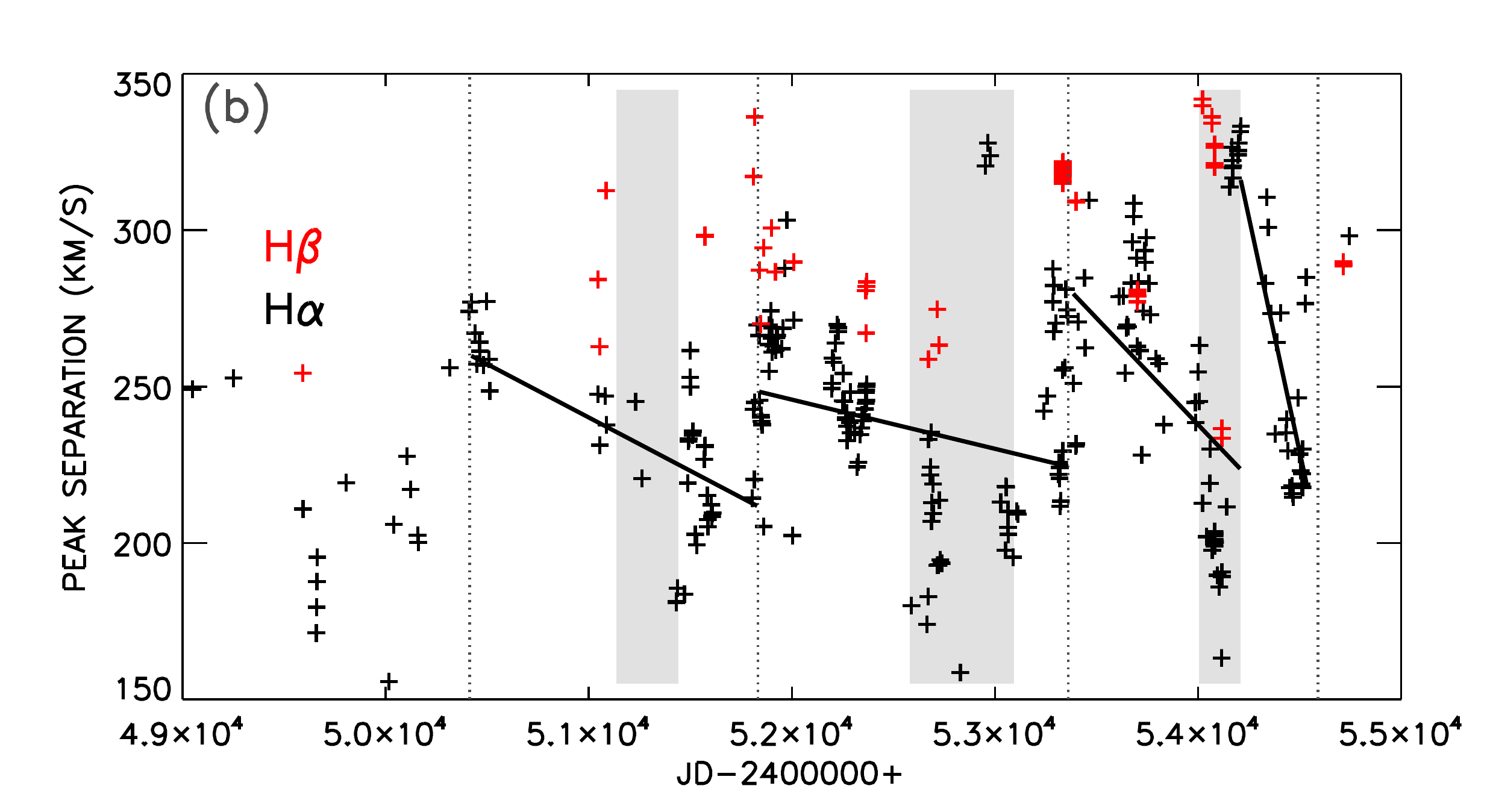} \\
        \includegraphics[trim=0cm 0cm 0.5cm 0.5cm, clip=true, angle=90, width=0.41\textwidth]{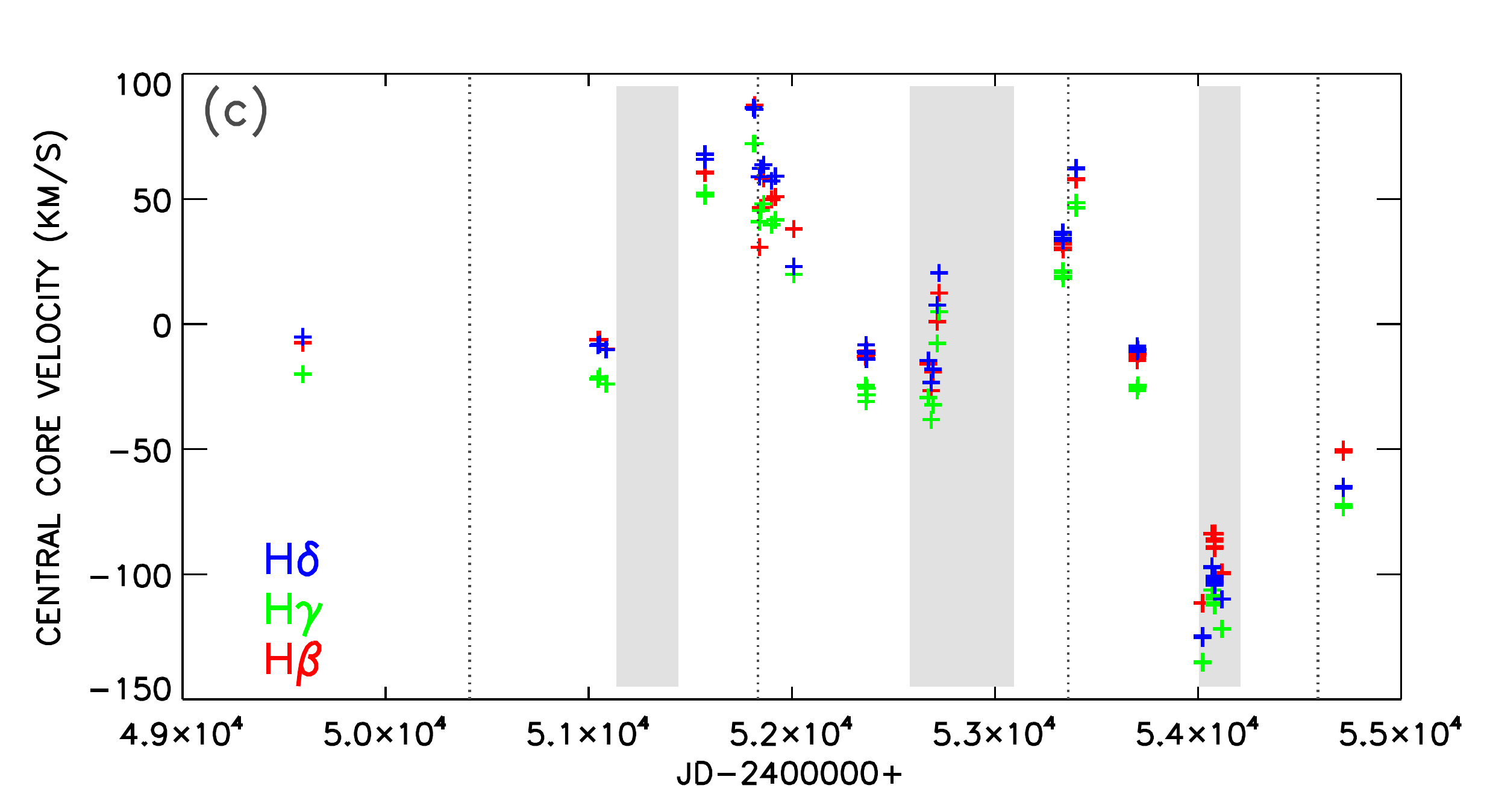} \\
        \includegraphics[trim=0cm 0cm 0.5cm 0.5cm, clip=true, angle=90, width=0.41\textwidth]{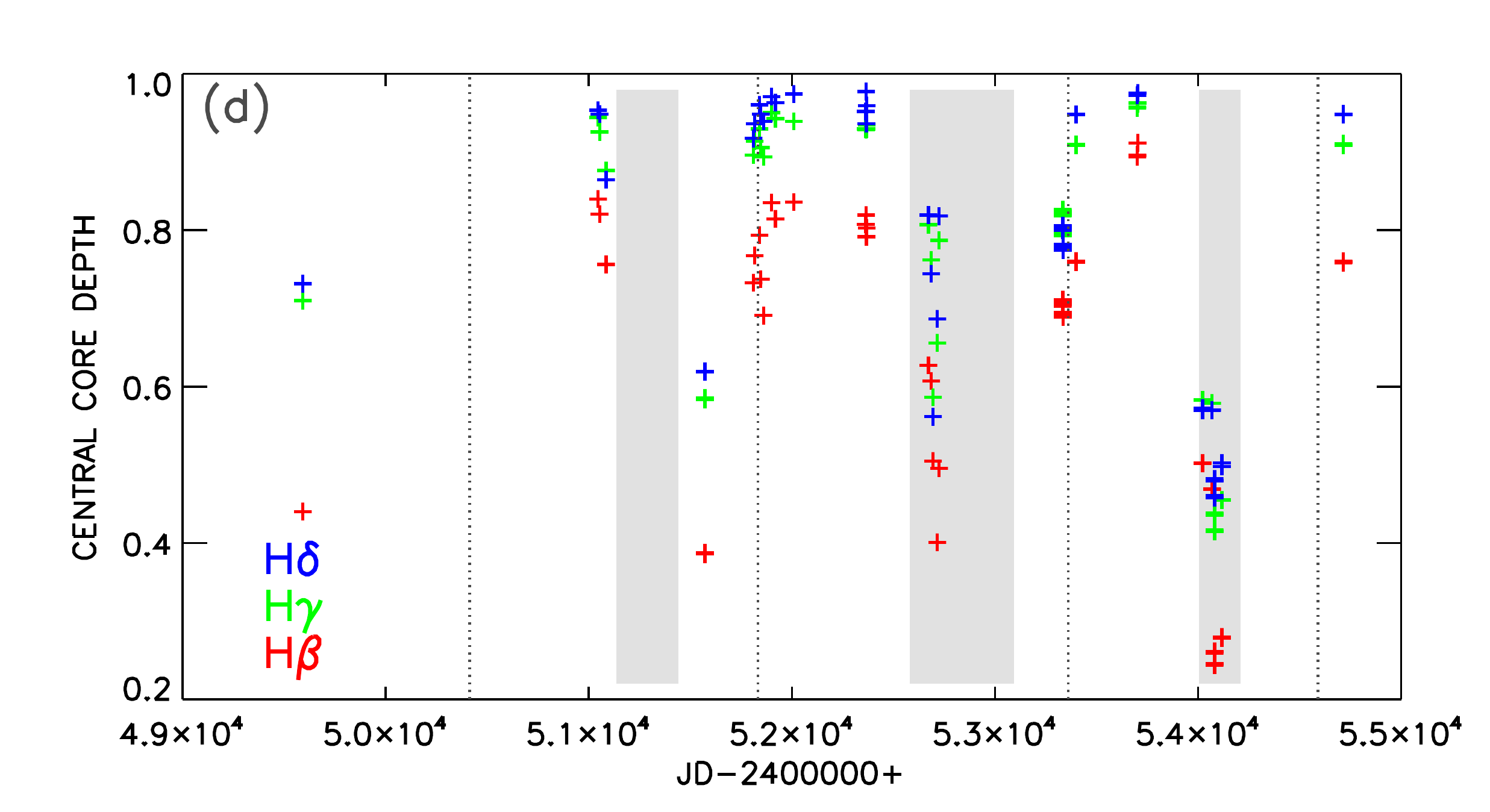} \\
        \includegraphics[trim=0cm 0cm 0.5cm 0.5cm, clip=true, angle=90, width=0.41\textwidth]{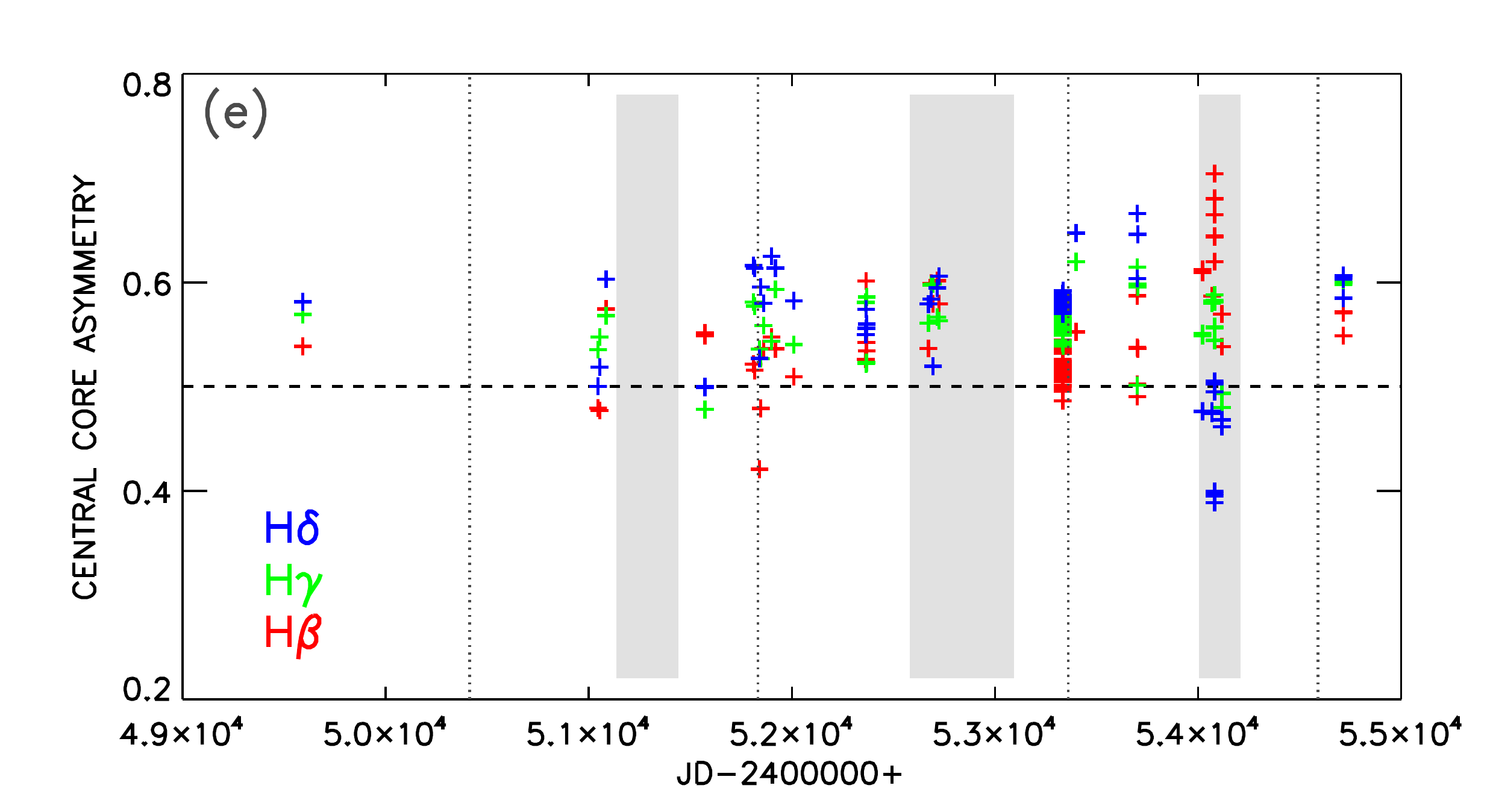} \\
        \caption{\emph{From top to bottom}: $V/R$ variations and peak separation (H$\alpha$, H$\beta$), position,
        depth, and asymmetry of the central absorption (H$\beta$, H$\gamma$, H$\delta$) as a function of time,
        expressed in Julian Days. The $V/R$ maxima are indicated by vertical dotted lines. The regions  in gray
        mark the triple-peak phases. Black, red, blue, and green symbols correspond respectively to H$\alpha$,
        H$\beta$, H$\gamma$, and H$\delta$ measurements. The black solid lines in panel (b) are linear regressions
        applied to H$\alpha$ peak separation.}
\label{fig3}
\end{figure}
\newpage

\section{Observed spectroscopic variations}
        \subsection{Emission line features}

The measured $V/R$ variations were presented in Fig.~1 (upper panel) and repeated in Fig.~3-a as reference
for the other observables discussed below. Only H$\alpha$ (black) and H$\beta$ (red) are plotted as only they
exhibit emission peaks pronounced enough to allow  meaningful $V/R$ values to be measured for entire cycles. 

A secondary $V/R$ maximum, whose position coincides with the triple-peak phase, can be identified for H$\alpha$
in Cycle~II and more marginally in Cycle~III. Unfortunately, the lack of measurements during these periods does not
allow a secondary $V/R$ maximum for H$\beta$ to be identified. The $V/R$ pattern observed in H$\beta$ is of smaller
amplitude and occurs slightly earlier than that of the H$\alpha$ line. Time lags between the $V/R$
variations of different emission lines (Balmer, Brackett, even \ion{He}{i} and heavier elements) were highlighted in
ZT\,I and \citet{wisniewski07}. Similar features were also found in EW\,Lac \citep{kogure84, mon13}. Since the
emission lines probe different radial zones of the unperturbed circumstellar disk, the time lag between their respective
$V/R$ cycles is evidence for the existence of a spiral density wave and, as will be shown in Sect.~4, an excellent
diagnostic tool for the global-disk oscillation model.

Despite the large scatter in the measurements, a saw-tooth pattern can be recognized for the peak separation (Fig.~3-b,
black solid lines). This is especially visible in Cycles~I and II, where the variation seems to follow the $V/R$ variation, with a
possible phase shift. In Cycle~III, because of the even more extreme scatter during the triple-peak phase, the distribution
looks like a double saw-tooth, with the second peak exhibiting a maximum separation larger than the first.

It was mentioned previously that the large-scale $V/R$ variations vanished after 2010. Some hints of a weakening density
perturbation are found in Figs.~3-a and 3-b. The first piece of evidence is the decreasing amplitude, from cycle to cycle, of the $V/R$
variations. The second  is the increasing peak separation at the $V/R$ maximum; the larger velocity difference
between the two peaks is the likely consequence of a disk progressively moving toward a lower density state by which
the one-armed oscillation moves out of resonance.

        \subsection{Shell line features}

The violet and red emission peaks are not the only ones to experience velocity shifts; the central shell component
exhibits oscillations of its radial velocity apparently synchronized with the $V/R$ variations (Fig.~3-c). When the
emission peaks are the farthest apart (\emph{\emph{i.e.}}, at the $V/R$ maximum), the central core reaches its maximum
redshift (v$_{rad}$ $\approx$\,$+80$\,km\,s$^{-1}$ and $\approx$\,$+60$\,km\,s$^{-1}$ at the first and second $V/R$
maximum,  respectively). Afterward, as the emission peaks move closer to each other and the $V/R$ ratio decreases,
the central core progressively shifts blueward. It reaches its maximum blueshift at the $V/R$ minimum, before shifting
back toward red. An extremely large blueshift (down to $\approx$\,$-100$\,km\,s$^{-1}$) occurs during the triple-peak
phase of Cycle~III. One can wonder whether these velocity shifts are not just artifacts resulting from the filling-in
effect of the lines on their blue side (at $V/R$ maximum) or on their red side (at $V/R$ minimum). This might be conceivable
for lines experiencing very strong $V/R$ variations, like H$\alpha$, but less obvious for H$\gamma$ and H$\delta$ in
which the filling-in effect is much weaker. In addition, the similar velocity shifts that are observed in lines from heavier
elements that do not experience any $V/R$ variations, like SiII\,$6347$ and HeI\,$6678$ (see Fig.~3 of \citealt{stefl09}),
favors the scenario in which the central absorption does experience a shift from its rest velocity, thus confirming that the
$V/R$ phenomenon is not only a matter of density structure, but also that of associated velocity field.

All the shell lines see their depth (Fig.~3-d) dropping very quickly during the $V/R$ maximum before going back rapidly
to their usual level. They all exhibit exceptionally shallow profiles during the triple-peak phase of Cycle~III, suggesting that
the origin of the triple peak may actually be an emission phenomenon filling in the central absorption, instead of an extra
absorption as hypothesized by ZT\,I. 

We can barely distinguish an overall consistent cyclic behavior for the shell line asymmetry (Fig.~3-e). The mean asymmetry
parameter is  0.545 for H$\beta$ (red), 0.562 for H$\gamma$ (blue), and 0.565 for H$\delta$ (green), which  indicates that
all the lines are on average slightly broader on their blue side. This blueward asymmetry seems to increase progressively for
higher members of the Balmer series; however, the difference from one line to another is extremely small. Intriguingly, the
asymmetry of H$\beta$ and H$\gamma$ are strongly and differently affected during the triple-peak phase of Cycle~III, the former
becoming broader on its blue side and the latter on its red side; H$\delta$ stays surprisingly unperturbed during that period.

We conclude that the depth and position of the shell absorption offer promising diagnostic potential for the $V/R$ phenomenon,
whereas the shell-line asymmetry does not, as it seemingly does not correlate with the $V/R$ cycle. This last statement should
nonetheless be moderated by the fact that shell profiles are, on average, broader on their blue edge, which should be
accounted for while testing our models as it is another hint of the spiral pattern.

\section{Modeling the data}
        \subsection{Tools and method}

The computer code HDUST \citep{carciofi06} was used to model the observed features described
above. It is designed to solve the coupled problems of radiative transfer, radiative equilibrium, and
statistical equilibrium in a non-local thermodynamical equilibrium (NLTE) regime. A Monte Carlo
approach is adopted: once emitted by the photosphere, the path of each photon across the stellar
envelope is followed. The radiative equilibrium is ensured because the absorbed photons are
reemitted with a direction and a wavelength determined by the local conditions. Both continuum
and line processes are included in order to compute the opacity and the emissivity of the gas.

The highly variable nature of $\zeta$\,Tauri requires a methodic approach and the combination of a
disk oscillation model with HDUST. The modeling campaign from ZT\,II was performed in two steps.
The first step consisted in fitting the average observed properties (optical and infrared SEDs, continuum
polarization H$\alpha$ peak separation) in order to fix the stellar and average disk parameters. The
theoretical background describing the structure of the unperturbed disk (see Appendix~A) assumes
a steady-state viscous decretion disk in hydrostatic balance in the vertical direction and truncated by
the binary at the tidal radius (18.6 $R_{*}$ for $\zeta$\,Tauri). The second step adopted the 2D global-disk
oscillation model of \citet{okazaki97} in order to reproduce the time-dependent properties of $\zeta$\,Tauri:
on the unperturbed state was superposed a linear $m=1$ perturbation in the form of a normal mode
of frequency $\omega$.

In this paper we adopt the model proposed by \citet{ogilvie08} for describing the dynamical
structure of the disk, and combine it with HDUST for computing line profiles and other observables.
The model of \citet{ogilvie08} includes the hitherto ignored $z$ component (in cylindrical coordinates
$\varpi$, $\phi$, $z$) in the calculations. This 3D formalism requires the perturbed quantities to be expanded
in the $z$-direction by means of Hermite polynomials (see Appendix~B). A positive aspect of these 3D
models is that, in contrast to their 2D counterparts, they are able to produce prograde confined
perturbation modes, as expected on both theoretical \citep{papaloizou92, oktariani09} and observational
grounds \citep{telting94, vakili98}, without using any weak line force \citep{chen94}.

Since this new model predicts a  complex structure for the circumstellar disk, we adopted in this work
a \emph{\emph{2.5D}} approach, resulting from the integration of the 3D perturbed disk structure in the $z$ direction.
This simplified approach is a first step toward a planned full 3D treatment. It allows  a meaningful comparison
between the new theory and the previous 2D formalism. In practice, the method consists first in solving the
3D equations (B.19 to B.21) describing the perturbed disk, the solutions of which are expressed by equations
B.24 to B.27. Afterward, by adopting a null value for the second-order terms ($w_{2}$ and h${_2}$), the
vertical components are neglected and only  the zeroth-order perturbed quantities, which are
independent of the vertical coordinate, remain. As will be seen below, this 2.5D model already brings various changes
to the model structure and, consequently, in the resulting observables. It qualitatively improves our modeling
capacities, thus building a useful bridge between the previous 2D and the upcoming 3D models.

\begin{figure*}[!ht]
        \centering
        \includegraphics[trim=0.cm 0.5cm 0.cm 0.cm, clip=true, angle=90, scale=0.7]{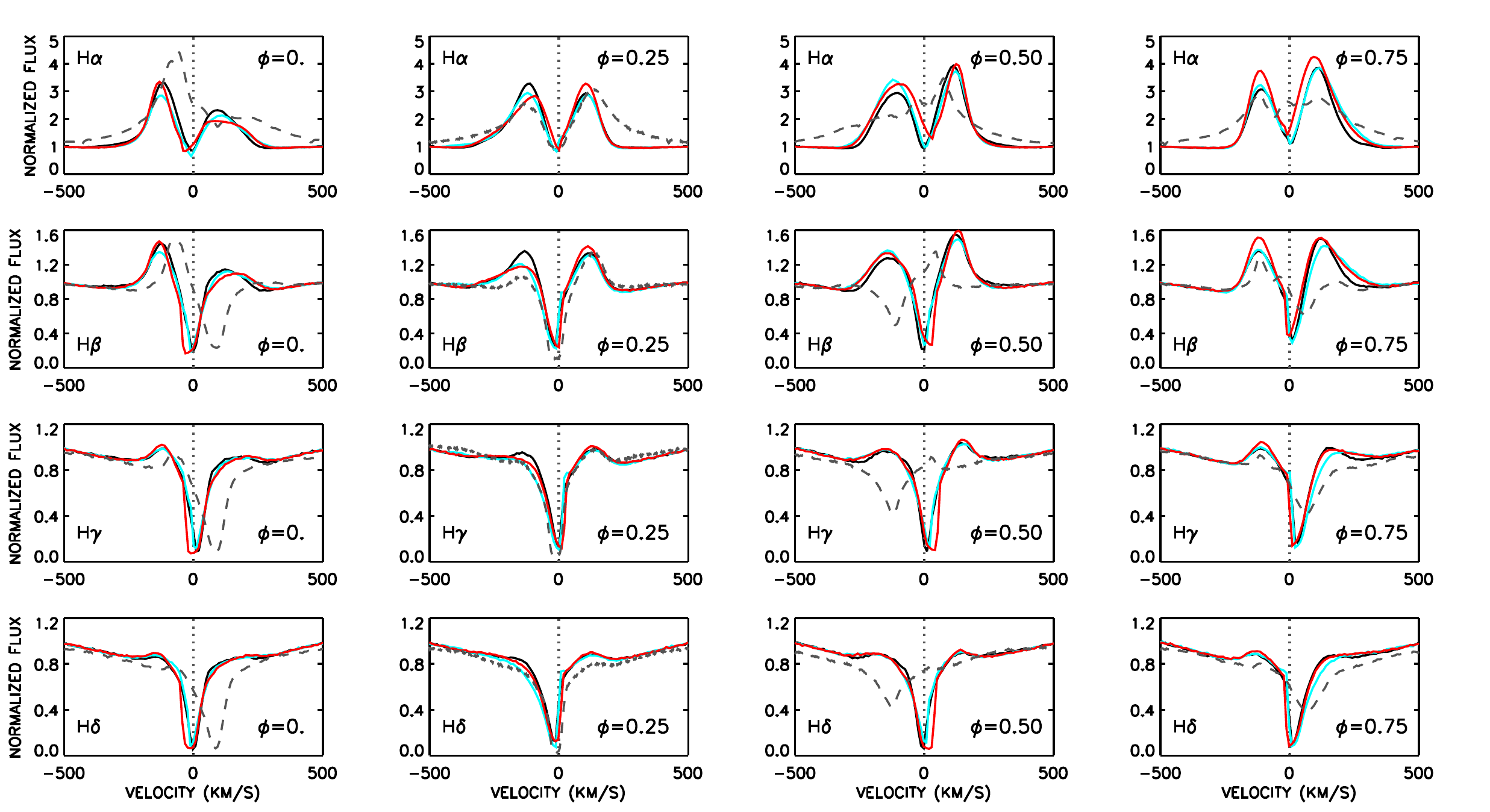}
        \caption{Optical Balmer lines as predicted by Model\,1 (blue), Model\,2 (cyan), and Model\,3 (red)
        at different phases of the $V/R$ cycle. From left to right are  shown the $V/R$ phases
        $\tau=$ 0., 0.25, 0.50, and 0.75; from top to bottom are  plotted H$\alpha$,
        H$\beta$, H$\gamma$, and H$\delta$. The vertical dotted line marks the rest
        velocity ($v_{rad} = 0$ km\,s$^{-1}$). The dashed curves are the observed profiles presented in
        Fig.~2 and are plotted here for comparison.}
\label{fig4}
\end{figure*}

        \subsection{Presentation of the models}

Our reference model (Model\,1) was computed with the 2D formalism and the stellar, orbital, and
geometric parameters constrained in ZT\,II (their Table~1). \citet{ogilvie08} noted that 3D
models are naturally more confined than their 2D counterparts. The degree of confinement of the perturbation
mode is defined by the viscosity parameter $\alpha$, first introduced by \citet{shakura73}. To allow a direct
comparison between the two formalisms, Model\,2 was run with the same parameters as Model\,1 and a
2.5D structure for the perturbed disk. In order to test the effect of the viscosity parameter, various 2.5D models
with different values of $\alpha$ were computed. The last model discussed below (Model\,3) was computed
with the same parameters as the other two  and $\alpha=0.8$ (instead of 0.4). It was chosen because it
provides a density wave confinement, materialized by the amplitude of its $V/R$ variations, comparable to
the one predicted by the reference model (Sect.~4.3 and Fig.~5). The respective disk structure of each model
will be discussed further in Sect.~5.1; their main differences are summarized in Table~1.

To make the comparison with our models easier, the observational data were converted in phase based
on the ephemerides measured in ZT\,I. The $V/R$ phase $\tau$ is usually defined such that $\tau = 0$ at
the $V/R$ maximum. To simulate the temporal variation the observables experience while the density wave
precesses around the star, for a given model we compute  the emerging spectrum for several azimuthal coordinates
$\phi$. The $V/R$ phase $\tau$ is related to the model phase $\phi$ by $$\tau= 1 - \frac{\phi - \phi_{0}}{2\pi},$$
where $\phi_{0}$ is the phase shift applied to each model in order to match the observational phase $\tau$
(see Table~1).

An illustration of the Balmer line profiles predicted by all three models is given in Fig.~4, at $V/R$ phases close
to those presented in Fig.~2. The features are qualitatively the same as those described in Sect.~3. The most
obvious difference is the absence of triple-peaked profiles in the models, suggesting that whatever the physical
process underlying this phenomenon is, it is neither included in nor predicted by our current models. This idea
is also supported by the very similar central absorption depths predicted by the models on phases $\phi = 0.25$
and $\phi = 0.75$, contrasting with the large differences observed in Fig.~2 between phases $\tau = 0.30$ and
$\tau = 0.81$. The same quantities presented in Fig.~1 (bottom panel) were measured on the model spectra, and
compared to the observational ones in Figs.~5 - 7.

\begin{table*}[ht]
\centering
\caption{Summary of the main parameters of our models}
\begin{tabular}{c c c c c c c c}
\hline
 & & & & \multicolumn{3}{c}{$V/R$ parameters} & \\
 \cline{5-8}
 & Stellar and disk parameters & Formalism & $\Phi_{0}$ & $k_{2}$ & Weak line force & Viscosity parameter $\alpha$ & \\
\hline
\hline
Model\,1 & ZT\,II & 2D  & 280$^{\circ}$ & 0.006 & 5.74$\times$10$^{-2}$($\varpi$/$R_{e}$)$^{0.1}$ & 0.4 & \\
Model\,2 & ZT\,II & 2.5D        & 35$^{\circ}$  & 0.008 & - & 0.4 & \\
Model\,3 & ZT\,II & 2.5D        & 55$^{\circ}$  & 0.009 & - & 0.8 & \\
\hline
\end{tabular}
\end{table*}

\begin{figure}[ht]
\centering
        \includegraphics[trim=0.cm 0cm 0.cm 0.cm, clip=true, width=0.48\textwidth]{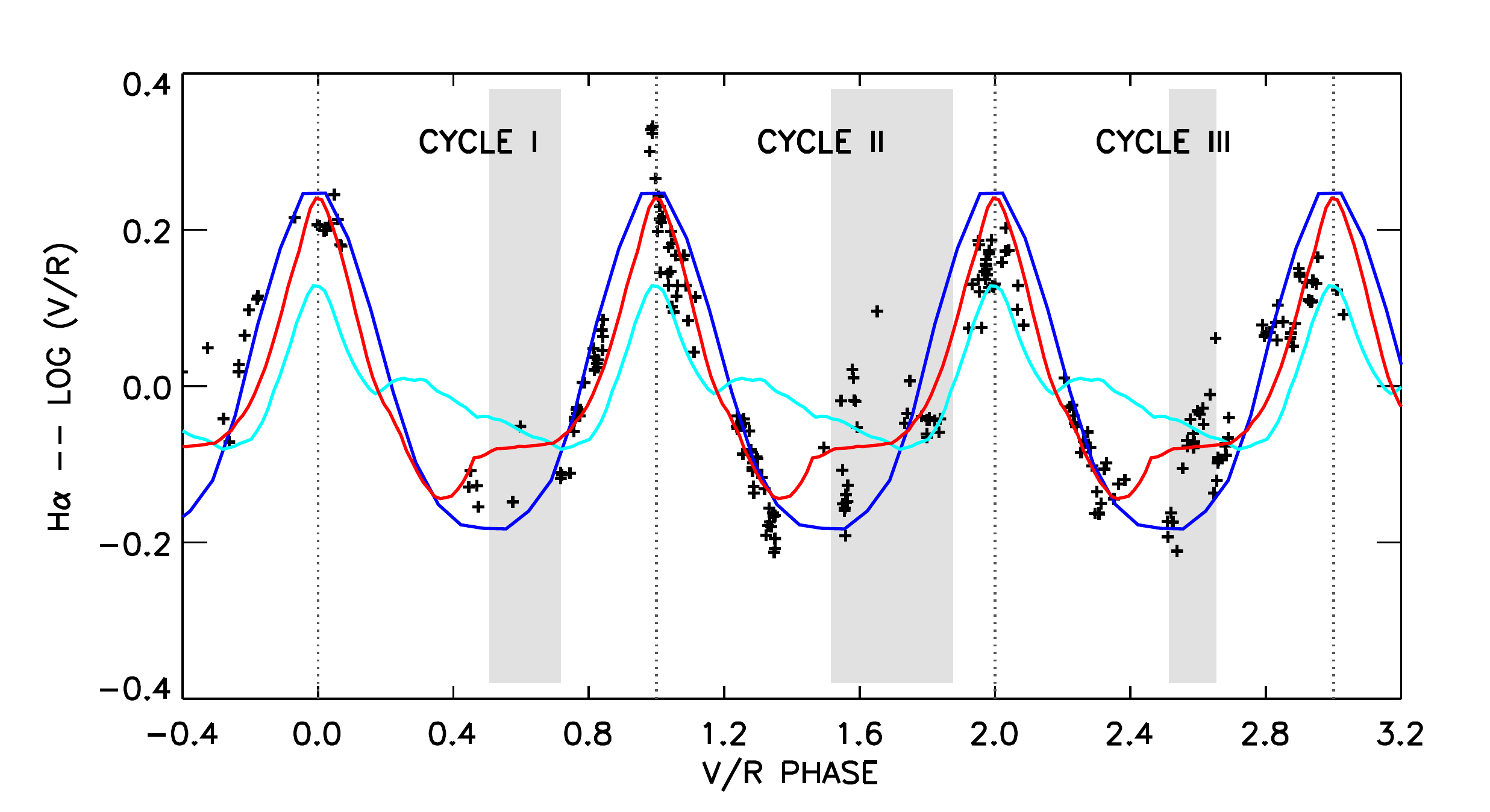} \\
        \includegraphics[trim=0.cm 0cm 0.cm 0.cm, clip=true, angle=90, width=0.48\textwidth]{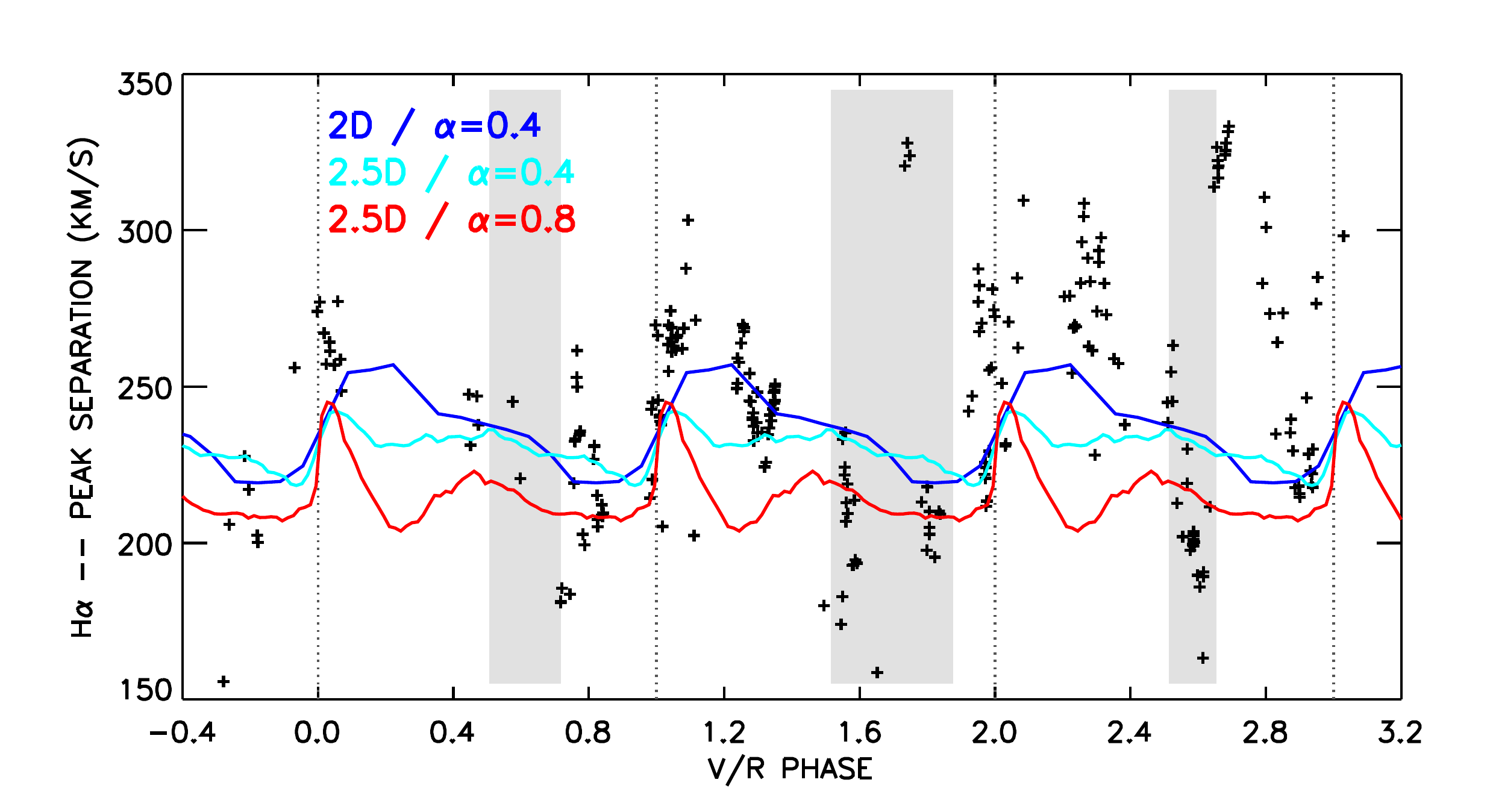} \\
        \caption{$V/R$ variations (upper panel) and peak separation (lower panel) of H$\alpha$.
        The observed points are the same as in Figs.~1 and ~3, but converted in phase to facilitate
        the comparison with our models. The solid blue line corresponds to Model\,1,  cyan  to
        Model\,2, and  red  to Model\,3. The regions in gray mark the triple-peak phases.}
\label{fig5}
\end{figure}

\begin{figure}[ht]
\centering
        \includegraphics[trim=0.cm 0cm 0.cm 0.cm, clip=true, angle=90, width=0.48\textwidth]{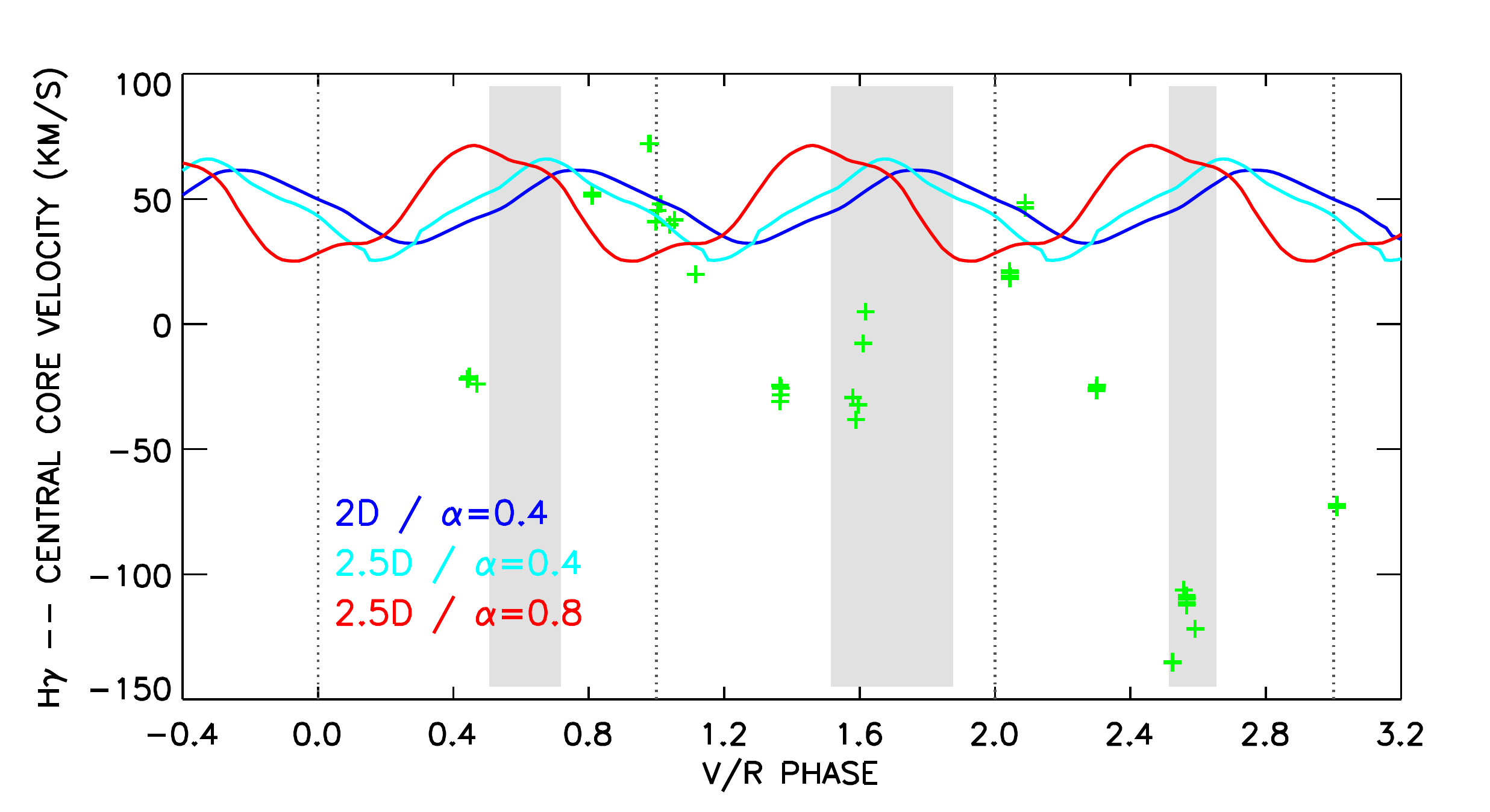} \\
        \includegraphics[trim=0.cm 0cm 0.cm 0.cm, clip=true, angle=90, width=0.48\textwidth]{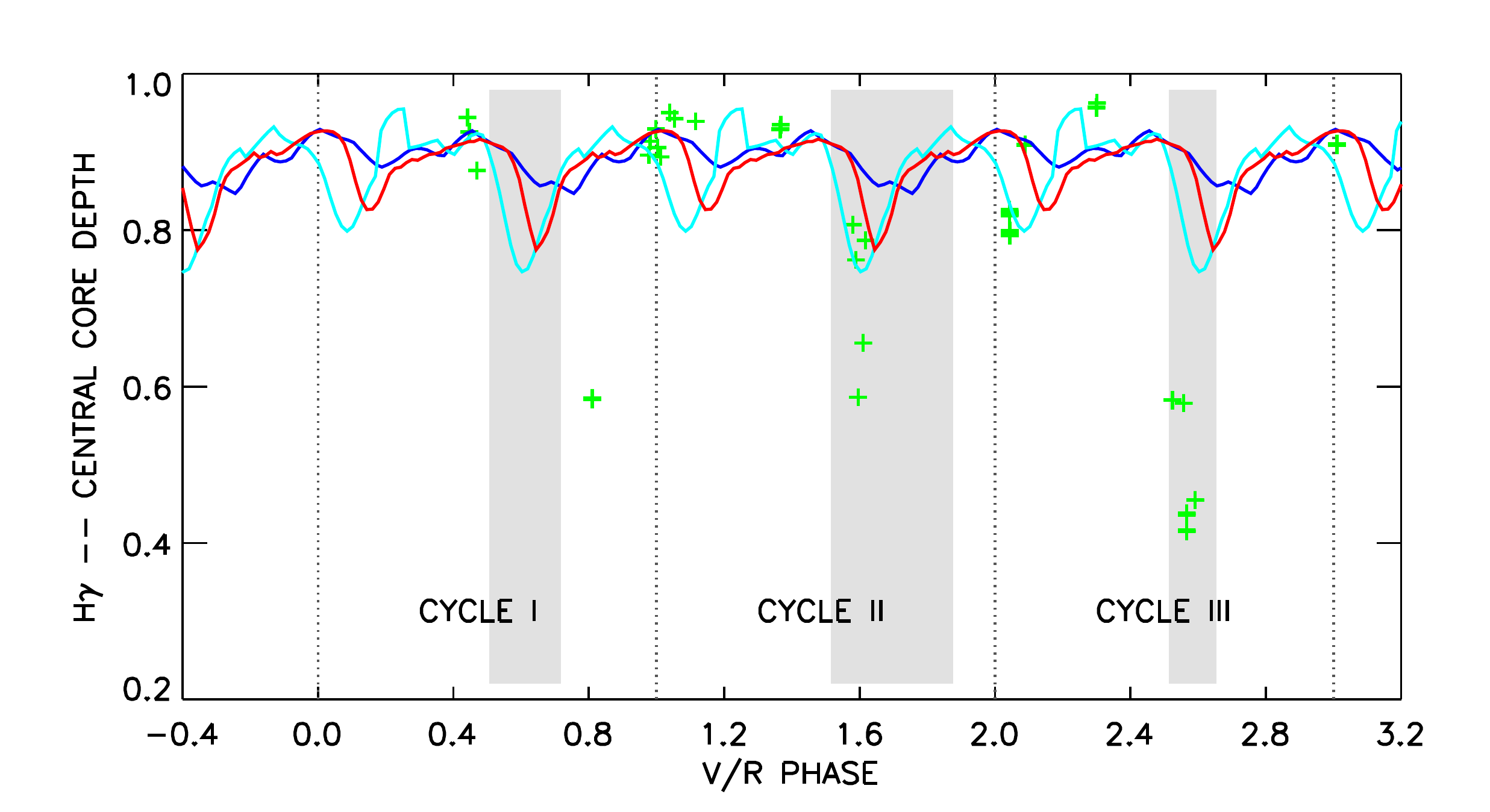} \\
        \includegraphics[trim=0.cm 0cm 0.cm 0.cm, clip=true, angle=90, width=0.48\textwidth]{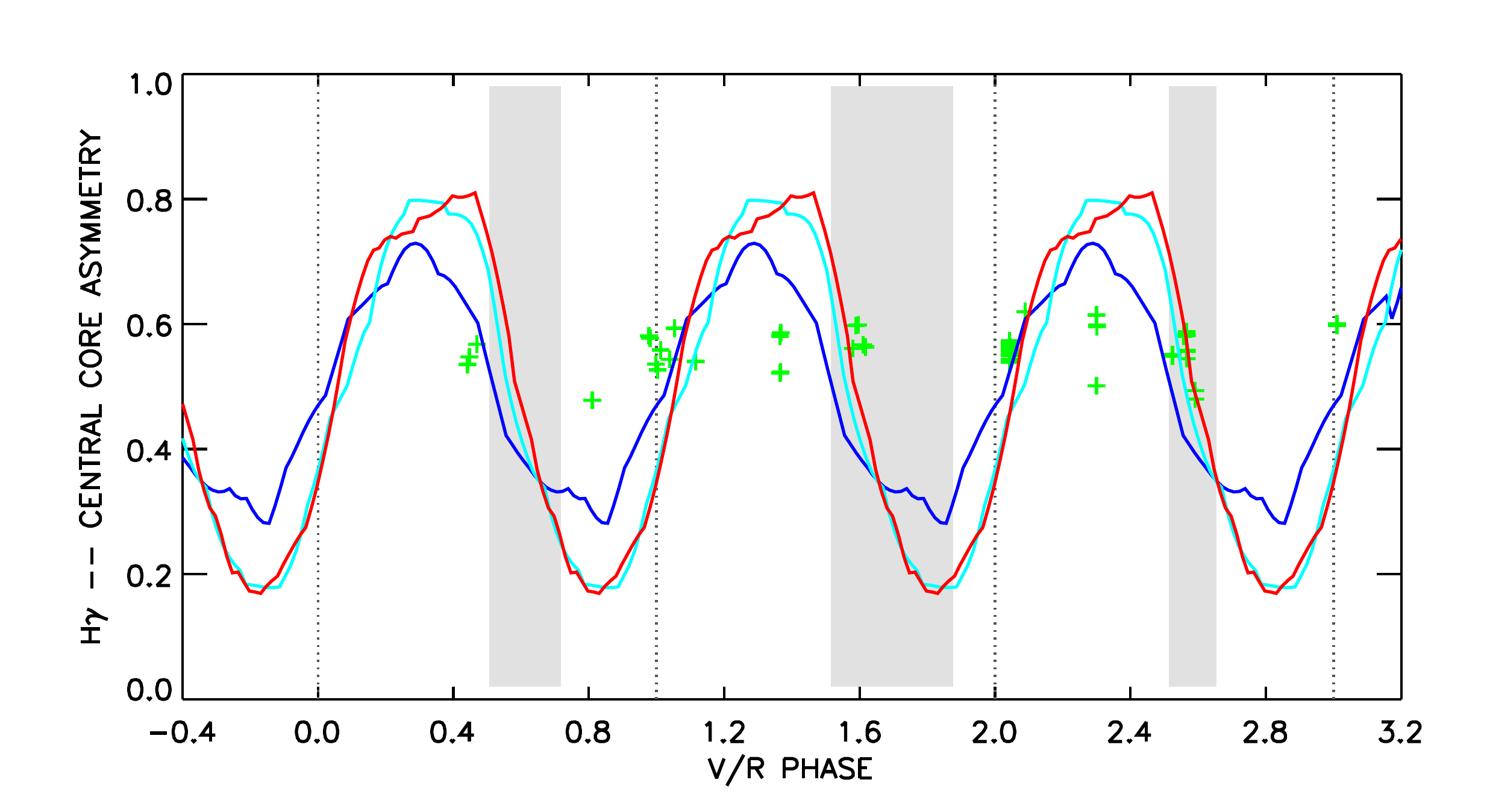} \\
        \caption{Shell line properties illustrated on H$\gamma$: central depth (top), central velocity (middle), and
        line asymmetry (bottom). The observational data are the same as in Fig.~3, but converted in phase to facilitate
        the comparison with our models. The solid blue line corresponds to Model\,1,  cyan  to Model\,2, and
         red  to Model\,3. The regions  in gray mark the triple-peak phases.}
\label{fig6}
\end{figure}

        \subsection{Emission line features}

The  properties of the emission lines are essential diagnostic tools for our models because  they probe the
physical conditions throughout the disk. The $V/R$ variations and peak separation predicted by
our models are plotted in Fig.~5 and compared to the observed ones. We chose H$\alpha$
to illustrate our results because the larger dataset allows  a better comparison. Nevertheless, all
the comments below apply identically to H$\beta$. Model 1 (from ZT\,II) provides a reasonable match
to the $V/R$ variations. Switching to the 2.5D formalism and keeping all the parameters fixed (Model\,2)
results in a more confined density wave, as demonstrated by a $V/R$ curve whose shape and amplitude
do not match the observations. This was fixed by increasing the viscosity parameter (Model\,3). The resulting
shape is quite different, however, as Model\,3 reaches the $V/R$ minimum sooner than Model\,1 and presents
a sort of plateau between $\tau \approx 0.5$ and $\tau \approx 0.8$. During the $V<R$ phases, the ability of
Model\,1 and Model\,3 to reproduce the observations does not seem qualitatively different. During the $V>R$
phases, Model\,3 predicts  a sharper $V/R$ profile, more in agreement with the observations. As expected from
the line profiles plotted in Fig.~4, the models are not able to explain the triple-peak phases (marked as gray-shaded
areas in this and the following plots for reference).

Reproducing the peak separation variations is more problematic. All the models exhibit
variations whose amplitude is smaller than what is observed. The 2D model (Model\,1) is the one that
matches best the saw-tooth pattern observed in Cycle~I and Cycle~II. The 2.5D models predict more
perturbed variations, especially Model\,3, which exhibits a secondary minimum that looks like the 
double saw-tooth pattern observed in Cycle~III but with a much smaller amplitude.

        \subsection{Shell line features}

While the emission lines sample the disk as a whole, the region probed by the shell lines is restrained
to the line of sight. The diagnostics they provide should not be neglected, as their properties
(absorption depth, central position, asymmetry) reflect the structure and dynamics of the disk in the
observer's direction. A comparison between our models and the observed shell-line properties is
proposed in Fig.~6, for H$\gamma$.

Looking first at the position of the central absorption (top panel), there is an obvious discrepancy between
the model predictions and the observations. First, the models are displaced in phase while for the same
value of $\phi_{0}$ they match  the $V/R$ phase of H$\alpha$ perfectly. Second, the average position
(similar for each model $\approx$ 47 km\,s$^{-1}$) is at a higher velocity  than the observed ($\approx$
11 km\,s$^{-1}$). Finally, the amplitude of the variations is clearly far too small in the models.

If we exclude the triple-peak phases, the models exhibit variations in the shell absorption depth whose
amplitude is compatible with the observations. The scarcity of the observational data does not allow 
the variation pattern predicted by the models to be validated.

The results for the line asymmetry (bottom panel) also highlight strong challenges for the models; they predict
large-amplitude (between 0.2 and 0.8) cyclic variations, while the observations show very little variation 
(amplitude $\approx$ 0.1) and no sign of cyclic behavior. Model\,3 predicts shell lines that are on average
slightly broader on their blue side (average asymmetry parameter $\approx 0.515$), as might be expected
from the results in Sect.~3.2, whereas Model\,1 predicts more symmetric shell line profiles (average asymmetry
parameter $\approx 0.505$) and Model\,2 shell lines slightly broader on their red side (average asymmetry
parameter $\approx 0.495$). The differences between the three models are marginal, however, considering
the typical error on the asymmetry parameter ($\pm 0.067$ for the spectral resolution of our models).

\begin{figure}[ht]
\centering
        \includegraphics[trim=0.cm 0.cm 0.cm 0.cm, clip=true, angle=90, width=0.48\textwidth]{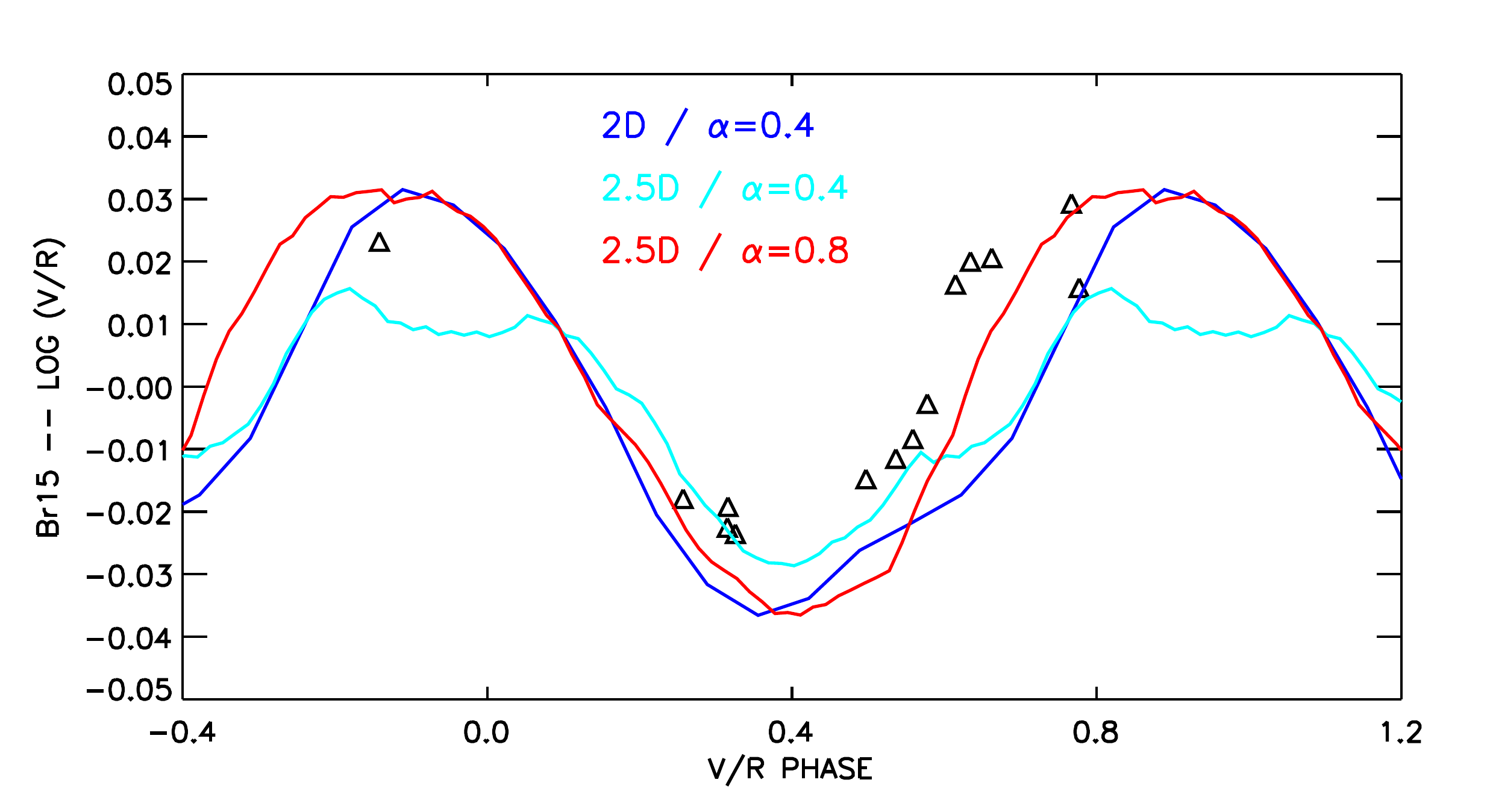} \\
        \includegraphics[trim=0.cm 0.cm 0.cm 0.cm, clip=true, angle=90, width=0.48\textwidth]{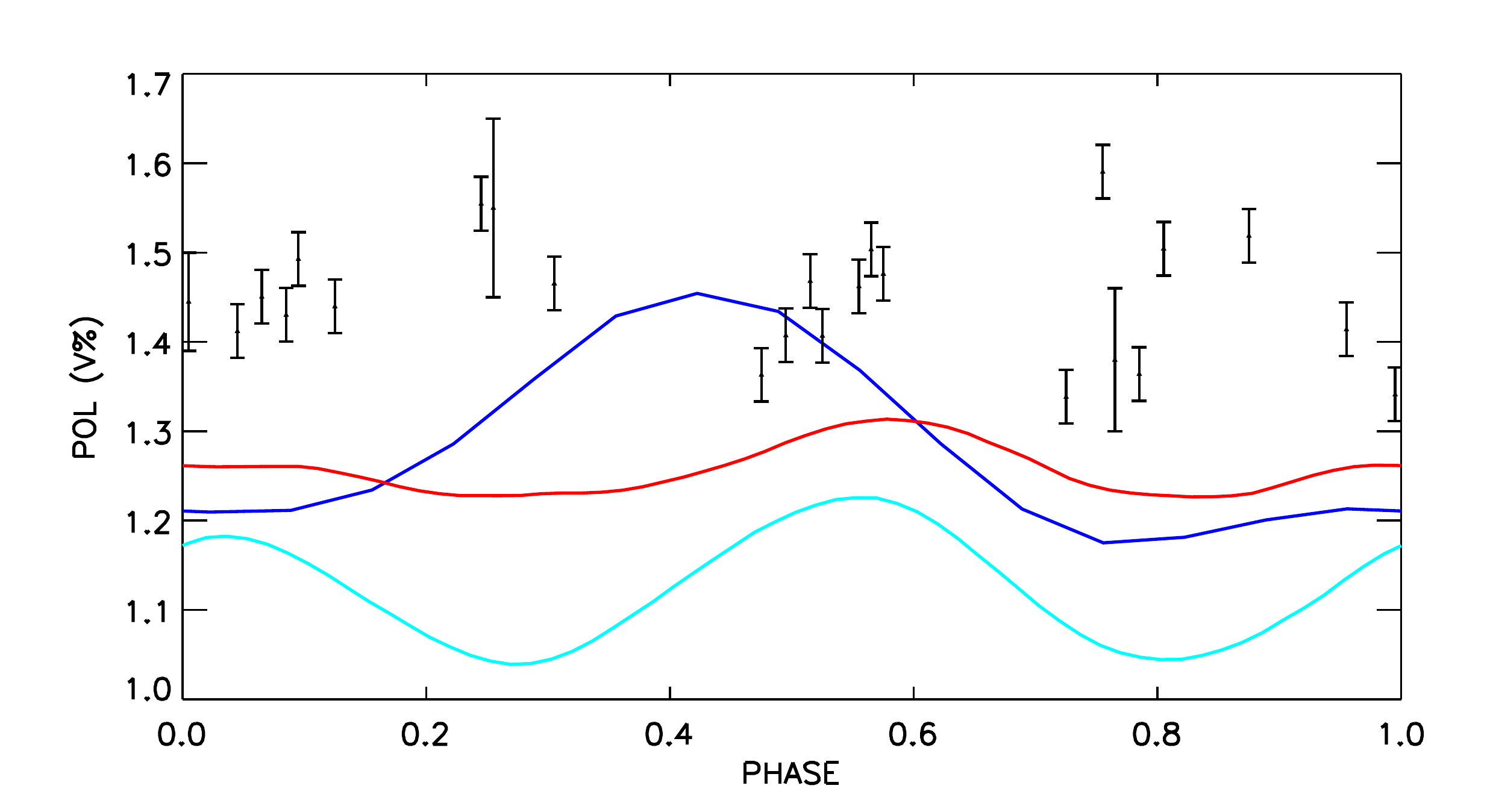} \\
        \caption{$V/R$ variations of \ion{Br}{15} (upper panel) and V-band polarization (lower panel). The solid blue
        line corresponds to Model\,1,  cyan  to Model\,2, and  red  to Model\,3.}
\label{fig6}
\end{figure}

Overall, these results highlight a limitation of our models in their current form. The 2D model was shown
by ZT~II to offer a good description of the spectro-interferometric data observed by VLTI/AMBER, which
are sensitive to the on-the-sky distribution of the disk intensity. This  and the good overall fit to the $V/R$
variations suggest that the models  reasonably reproduce the density distribution of gas across the disk.
However, the line features of $\zeta$ Tauri, considered in such detail for the first time, are influenced
not only by the density distribution, but also by the disk kinematics. In particular, the shell features  are
sensitive to the line-of-sight projected velocity of the disk. Therefore, the general failure of both the 2D
model by \citet{okazaki97} and \citet{papaloizou92} and the present 2.5D model based on the works of
\citet{ogilvie08} to reproduce them suggests that both models are predicting the wrong kinematic
structure for the disk.

\begin{figure*}[t]
        \centering
        \includegraphics[trim=0.5cm 0cm 8.5cm 0.cm, clip=true, scale=0.35]{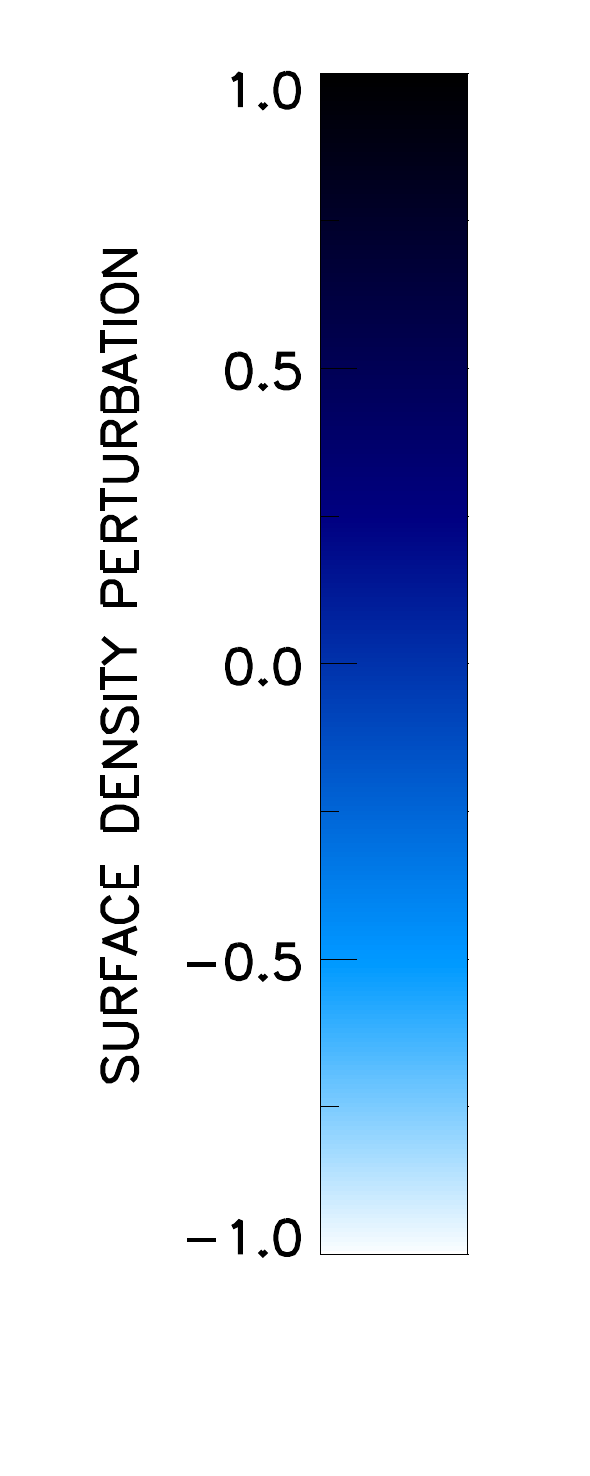}
        \includegraphics[scale=0.35]{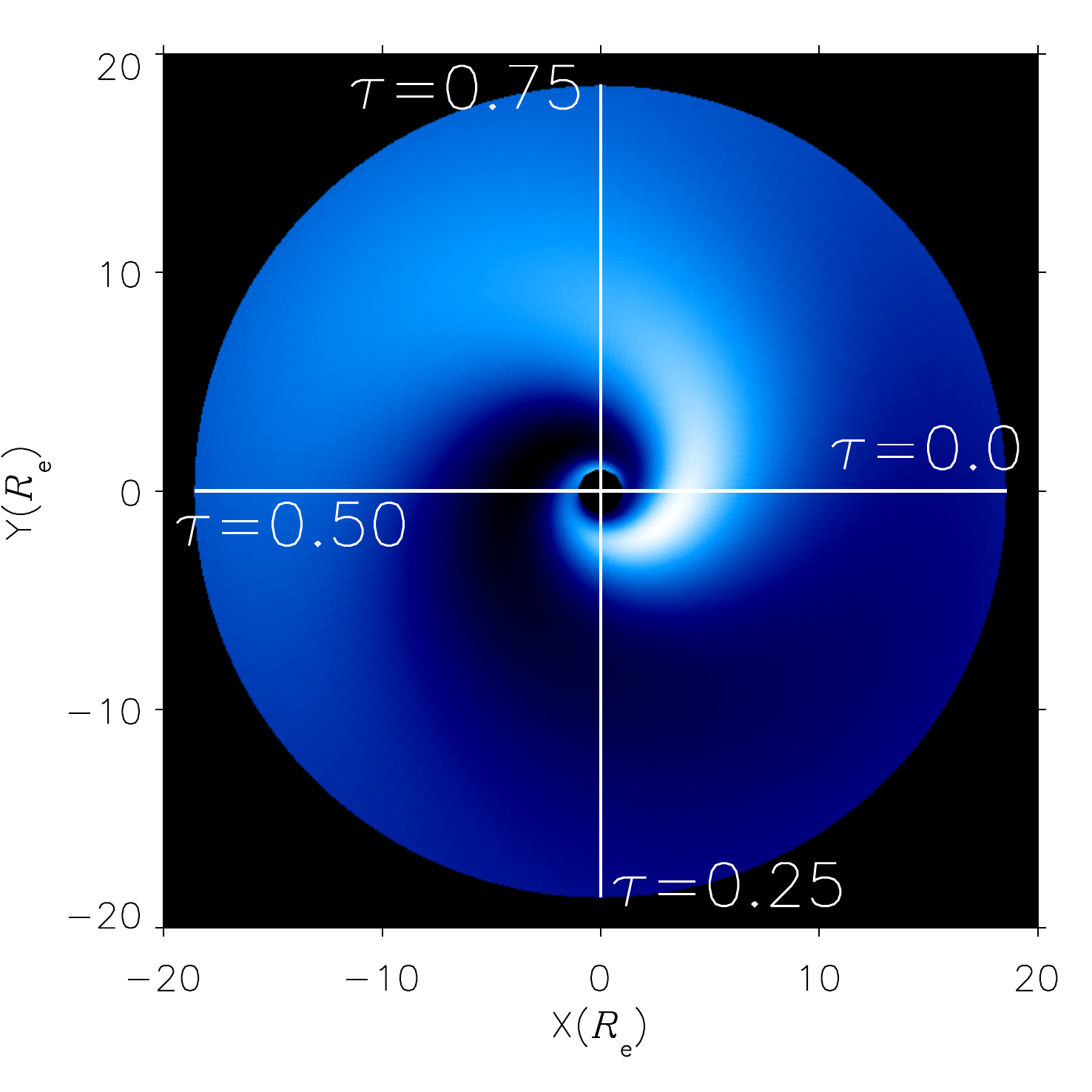}
        \includegraphics[scale=0.35]{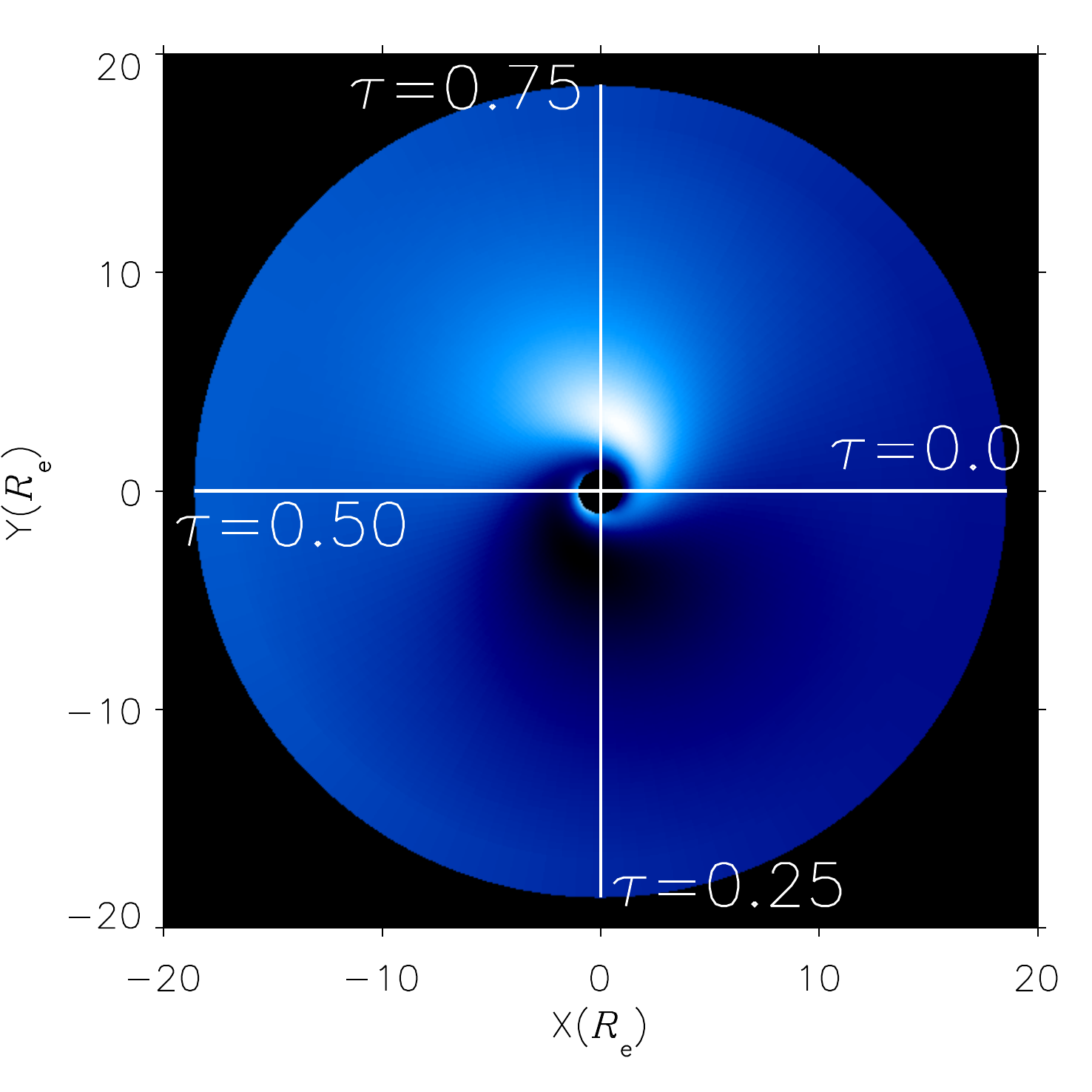}
        \includegraphics[scale=0.35]{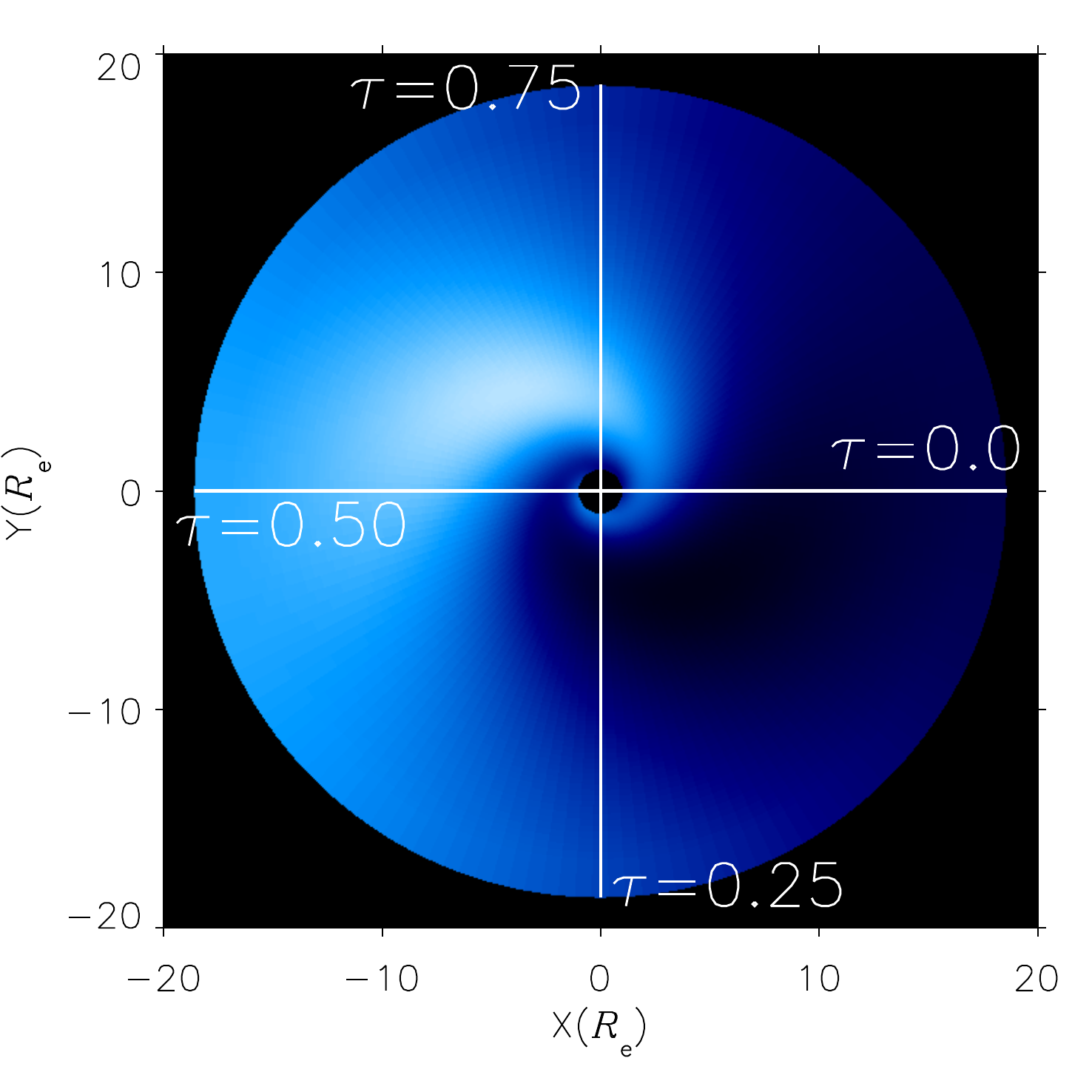}
        \caption{Density perturbation pattern as predicted by three different disk oscillation models. From left
        to right are presented a 2D model with $\alpha = 0.4$ (Model\,1), a 2.5D model with $\alpha = 0.4$
        (Model\,2), and a 2.5D model with $\alpha = 0.8$ (Model\,3). In these representations, the disk is seen
        from above and the x- and y-axes express the distance from the center of the star in stellar radii. The
        color scheme indicates the amplitude of the surface density perturbation ($\Sigma$'), normalized by
        the unperturbed value ($\Sigma$). Darker colors indicate over-densities and lighter color sub-densities.
        The line of sight toward the observer is marked in four directions, sampling the $V/R$ cycle in phase
        from $\tau = 0.$ (\emph{\emph{i.e.}}, maximum $V/R$) to $\tau = 0.75$.}
\label{fig8}
\end{figure*}

        \subsection{IR lines and polarization}

Two important issues were highlighted in ZT\,II concerning the 2D global-disk oscillation model.
The first  was its inability to reproduce the phase of the $V/R$ cycle of \ion{Br}{15}. The
model reproduced the observed amplitude, but had to be artificially shifted by 0.2 in phase in order to match
the observations. The second  was about the polarization: the model predicted variations whose
amplitude was much larger than the observed one. Since both the IR lines and the polarization originate
from the innermost regions of Be disks, it was concluded in ZT~II that the 2D model was unable to predict
the correct disk structure in these regions. This was supported by the work of \citet{ogilvie08}, who showed
how much stronger the confinement of the oscillation modes is when a 3D perturbation is adopted.

Figure~7 (top panel) compares the three aforementioned models with the observed \ion{Br}{15} $V/R$ cycle
(data from ZT~I). Models~1 and 3 both reproduce the amplitude of the variations, but Model\,3 clearly does
a better job at reproducing the $V/R$ phase. Model\,2 obviously presents the worse fit quality, as was the case
for the H$\alpha$ cycle. It was evoked previously that the time lag, or equivalently the phase shift, between
the $V/R$ maxima of different emission lines is an important diagnostic tool for the global-disk oscillation model.
The ability of the 2.5D formalism to  consistently fit the $V/R$ variations of H$\alpha$ and \ion{Br}{15}
represents a great step in our understanding of the phenomenon, as it puts some constraints on the
properties of the spiral arm perturbation, further discussed in Sect.~5.

We now consider the continuum polarization (Fig.~7, bottom panel). The observations show an average
level of 1.45$\%$ and a standard deviation of $\pm$0.07$\%$. Model\,2 is the model providing the smallest
polarization level (1.12$\%$ in average). Model\,1 and Model\,3 exhibit a higher level (1.28$\%$ and 1.26$\%$,
respectively) but remain below the observations. This result tends to favor less confined models, but we  see
in Sect.~5.2 that other factors can produce an average level comparable to the observed one. Regarding the
amplitude of the variations, Model\,2 matches  the observations better, while Model\,1 and Model\,3 predict
 too large large ($\pm$0.14$\%$) and too small variations ($\pm$0.05$\%$), respectively. To summarize, none
of these models provides a satisfactory fit to the polarization data, as they can either match the average level or the
amplitude, but not the two simultaneously. On a more qualitative note, the shape of the polarization variations
predicted by the 2.5D models present two maxima per $V/R$ cycle, as expected on observational grounds,
while the 2D formalism presents a centered, single-maximum pattern.

\begin{table*}[ht]
\centering
\caption{Summary of the main parameters for the models presented in Sect.~4 and Sect.~5 and their impact on the following
observables: amplitude of the $V/R$ variations $\Delta$\,$V/R$, average peak separation $<{\rm v_{peak}}>$, amplitude
of the peak separation variations $\Delta$\,v$_{peak}$, average polarization level $<Pol(V\%)>$, and amplitude of the polarization
variations $\Delta$\,Pol(V$\%$).}
\begin{tabular}{c c c c c c c c c c}
\hline
Model & H$_{0}$ (R$_{\odot}$) & $\rho_{0}$ (g\,cm$^{-3}$) & $i$ & $\alpha$ & $\Delta$\,$V/R$ (log) & $<{\rm v_{peak}}>$ (km\,s$^{-1}$) &
$\Delta$\,v$_{peak}$ (km\,s$^{-1}$) & $<{\rm Pol}(V\%)>$ & $\Delta$\,Pol(V$\%$) \\
\hline
Model\,1                & 0.208 & 5.90$\times$10$^{-11}$        & 85$^{\circ}$  & 0.4             & 0.495 & 218.5 & 24.1 & 1.284 & 0.279 \\
Model\,2                & 0.208 & 5.90$\times$10$^{-11}$        & 85$^{\circ}$  & 0.4             & 0.209 & 230.8 & 23.8 & 1.122 & 0.199 \\
Model\,3                & 0.208 & 5.90$\times$10$^{-11}$        & 85$^{\circ}$  & 0.8             & 0.384 & 215.3 & 41.2 & 1.256 & 0.098 \\
$\rho_{1}$      & 0.208 & 7.08$\times$10$^{-11}$        & 85$^{\circ}$  & 0.8             & 0.348 & 209.2 & 34.7 & 1.269 & 0.097 \\
$\rho_{2}$      & 0.208 & 8.85$\times$10$^{-11}$        & 85$^{\circ}$  & 0.8             & 0.314 & 202.7 & 22.9 & 1.291 & 0.119 \\
$\rho_{3}$      & 0.208 & 1.18$\times$10$^{-10}$        & 85$^{\circ}$  & 0.8             & 0.317 & 196.3 & 18.1 & 1.314 & 0.146 \\
i$_{1}$         & 0.208 & 5.90$\times$10$^{-11}$        & 83$^{\circ}$  & 0.8             & 0.319 & 209.6 & 27.3 & 1.357 & 0.114 \\
i$_{2}$         & 0.208 & 5.90$\times$10$^{-11}$        & 87$^{\circ}$  & 0.8             & 0.447 & 220.9 & 61.6 & 1.129 & 0.113 \\
H$_{1}$         & 0.222 & 5.90$\times$10$^{-11}$        & 85$^{\circ}$  & 0.8             & 0.397 & 216.8 & 44.3 & 1.304 & 0.107 \\
H$_{2}$         & 0.236 & 5.90$\times$10$^{-11}$        & 85$^{\circ}$  & 0.8             & 0.416 & 218.9 & 46.0 & 1.344 & 0.112 \\
H$_{3}$         & 0.249 & 5.90$\times$10$^{-11}$        & 85$^{\circ}$  & 0.8             & 0.426 & 220.4 & 48.1 & 1.390 & 0.115 \\
\hline
\end{tabular}
\end{table*}

\section{Discussion}
        \subsection{Structure of the solution}

In the previous sections, we showed that switching from the 2D to the 2.5D formalism and
increasing the viscosity parameter, $\alpha$, imply large modifications on the observables
computed by our models. 
In this section we discuss how these modifications reflect important changes in the model disk
structure, especially regarding the shape of the spiral density wave. 

An illustration of these structural differences is given in Fig.~8, where  the
disk density perturbation pattern is represented for the three models detailed previously. The shading
indicates the amplitude and sign of the perturbation, from density deficit (lighter colors)
to density enhancements (darker colors). A simple visual comparison between Model\,1
and Model\,2 (left and center panels, respectively) shows qualitatively how the strength of the
confinement of the density wave changes while moving from the 2D to the 2.5D formalism: in
the former, the spiral arms extend up to $\approx10\,R_{e}$ while in the latter they are
restrained within $\approx5\,R_{e}$.

It is only by adopting a high value of $\alpha$ that the 2.5D formalism produces density
perturbations able to spread at larger radii (Model\,3,  right panel). That parameter was
introduced by \citet{shakura73} to characterize the efficiency of the angular momentum
transport in a turbulent, differentially rotating medium. Higher values of $\alpha$ imply a
more efficient diffusion of the density wave, explaining why the perturbation spread
farther in Model\,3. The results of \citet{shakura73} suggested that $\alpha$ cannot be
greater than 1 in the case of turbulent viscosity, so adopting a value as large as 0.8 in our
models can almost be seen as a limit case (see, e.g., the recent determination of
$\alpha=1.0\pm0.2$ for the disk of 28\,CMa made by \citealt{carciofi12}).

The spiral arm perturbation can be characterized not only by its extent/confinement, but
also by how tight it is wound around the star. On that score, it can be seen in Fig.~8 that
the 2.5D models exhibit more coiled spiral patterns than the 2D model. Keeping in mind that ZT\,II
defined $\phi_{0}$ as the azimuth of the minimum of the density perturbation pattern at the
base of the disk, the different coil from one model to another is what explains the large phase
shift seen between Model\,1 ($\phi_{0} = 280^{\circ}$) and Models~2 and 3
($\phi_{0} = 35^{\circ}$ and $\phi_{0} = 55^{\circ}$, respectively). In addition, the tighter coil
of the spiral arm of Model\,3, compared to Model\,1, explains why the first does a better job
at fitting the $V/R$ phase of \ion{Br}{15} than the second.

        \subsection{Diagnostic potential}

\begin{figure*}
        \centering
        \includegraphics[scale=0.7]{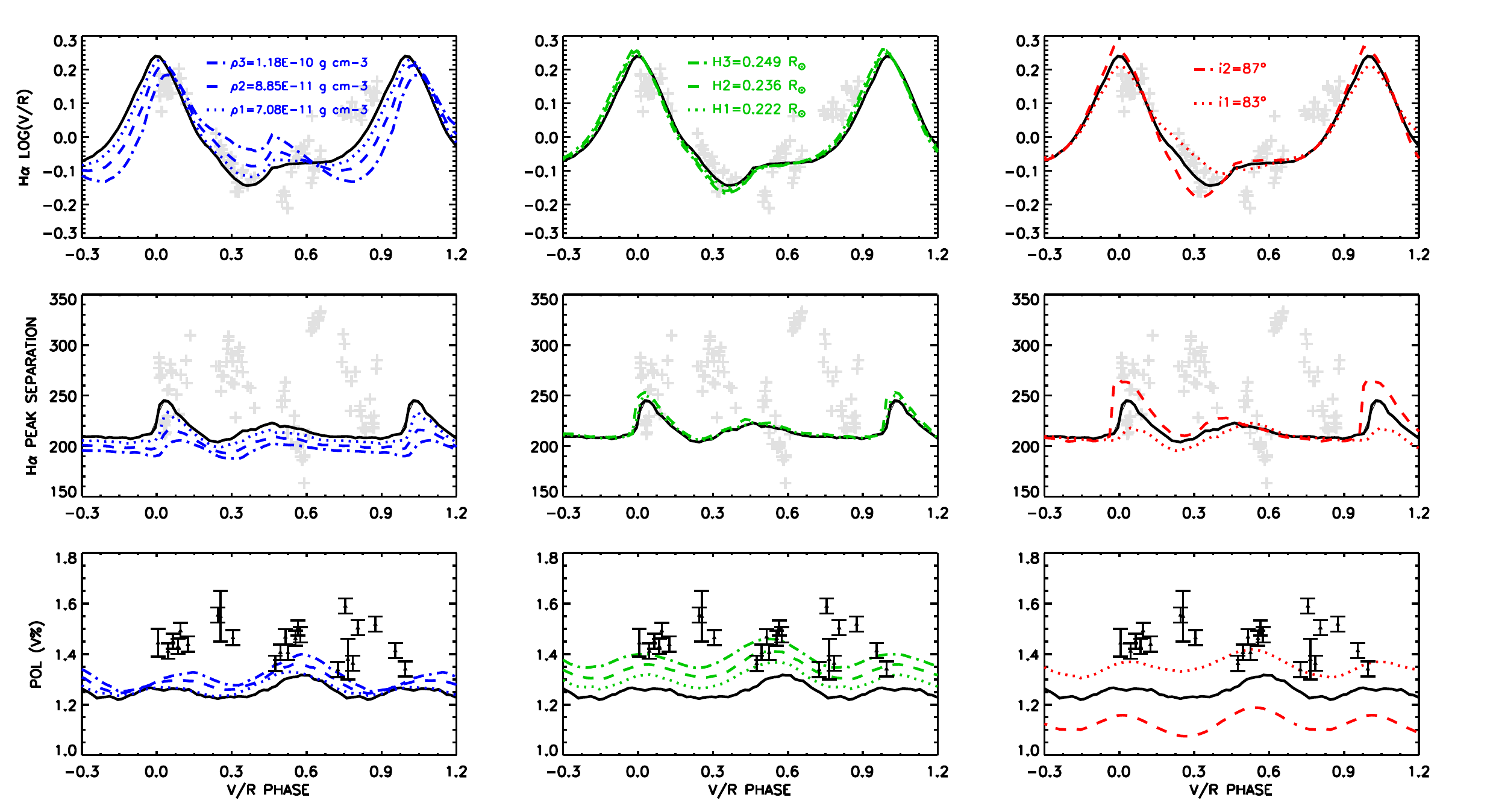}
        \caption{Testing the effect of the disk base density (left column,  blue), vertical
        scale height (middle column,  green), and inclination angle (right column,  red) on the $V/R$ variations (top and center row) and V band polarization (bottom
        row) predicted by HDUST models. The black solid line represents Model\,3, taken
        here as a reference for our tests. The model parameters are summarized in Table~2.
        For clarity, we limited the $V/R$ observational sample at Cycle~I only and plotted it
        in light gray.}
\label{fig9}
\end{figure*}

Overall, the previous results indicate that the 2.5D formalism provides a better representation
of the global-disk oscillation phenomenon, at least regarding the intensity diagnostics, such as
the $V/R$ curves of H$\alpha$ and \ion{Br}{15}. It remains less conclusive regarding the
kinematics diagnostics, such as the shell features.

Various model parameters determined in ZT\,II were kept fixed for the analysis presented above.
Among these parameters, some may strongly affect the on-the-sky projected disk density, namely
the base density of the disk, its vertical scale height, and its inclination angle with respect to the
observer. The impact of these three quantities on the observables are now studied to test whether
the values determined in ZT\,II remain the best ones in the 2.5D formalism framework.

Model 3 (2.5D with $\alpha=0.8$) will be taken as  reference here; starting from that model, we
made the (unperturbed) base density $\rho_{0}$, the scale height $H_{0}$ and the inclination
angle $i$ vary. The model parameters are compiled in Table~2 and the results pictured in Fig.~9
for three observables: the H$\alpha$ $V/R$ variations (top row),  peak separation (middle row),
and the continuum polarization (bottom row).

We consider first the unperturbed base density (left column, blue), \emph{\emph{i.e.}}, the density
(expressed in g\,cm$^{-3}$) in the equatorial plane of the disk and at its innermost radius. The models
presented in Sect.~4 were computed with $\rho_{0}$ =  7.9$\times$10$^{-11}$ g\,cm$^{-3}$; we
increased that value by  20, 50, and 100\%. Doing so tends to progressively reduce the
amplitude of both $V/R$ variation and peak separation, hence degrading the fit to these two observables.
On the other hand, higher densities tend to match  the polarimetric data slightly better,  although without
reaching the observed level. As highlighted by \citet[their Fig.~4]{haubois14}, the relation between
polarization and density is not straightforward as it also depends strongly on the effective temperature
of the central star. For tenuous disks, the polarization level increases linearly with the density, whatever
the effective temperature is. For denser disks, the growing contribution of multiple-scattering and
pre-scattering absorption leads to a non-linear variation of the polarization with base density. Zeta Tauri
seems to lie in a regime where raising its base density by a large amount does not affect its polarization
level significantly.

In the isothermal viscous decretion disk model \citep{carciofi06}, the radial density fall-off is a power law
with a slope of $-3.5$ and the vertical density structure is a Gaussian with a scale height given by
$$H(r) = H_{0} \left(\frac{r}{R_{e}}\right)^{\beta},$$ where, $$H_{0} = c_{s}V_{\rm crit}^{-1}R_{e}$$ is
the scale height at the base of the disk, $c_{s}$ the sound speed, $V_{\rm crit}$ the critical velocity,
and $\beta=1.5$ for isothermal disks. In ZT\,II, $H_{0}$ was fixed to 0.208\,$R_{\odot}$ by adopting
a gas temperature of about 70\% of the stellar effective temperature on the equatorial plane. For our
tests, we increased $H_{0}$ up to 0.249\,$R_{\odot}$, corresponding to a gas temperature equal to
the effective temperature. The effect on the observables can be summarized as follows: \emph{(i)} the
$V/R$ curve appears little affected; \emph{(ii)} the maximum peak separation slightly increases, but still
remains well below the observed values; \emph{(iii)} the average polarization increases to a level
comparable to the observations while the amplitude of the variations remains almost unchanged,
and thus is a bit too small.

The inclination angle was determined in ZT\,II through the fits to SED and continuum polarization.
It was shown at the time that a difference as small as 2$^{\circ}$  greatly affects the polarization.
This is confirmed in Fig.~9 (right column, red): the average polarization level decreases
($resp.$ increases) significantly for slightly higher ($resp.$ lower) inclination, with respect to the
reference model. A lower inclination angle ($i=83^{\circ}$, dotted line) seems more consistent with
the observed polarization level, but it slightly degrades the fit to both the $V/R$ variations and peak
separation. Instead, a higher inclination angle ($i=87^{\circ}$, dashed line) provides larger
values of the maximum peak separation, almost comparable to the observed maximum, but then the
polarization level becomes far too low.

In summary, it appears possible  to improve the match with the polarization data by tuning the three
aforementioned quantities, but this implies degrading the fit to the observed $V/R$ variations. It is also
possible to improve the match with the observed peak separation, at the cost of making the match with
the polarization data even worse. From those considerations, the base density and inclination angle
determined in ZT\,II  remain the best parameters to fit the $V/R$ quantities. This was not unexpected
since they were determined independently and without any former consideration for the disk oscillation
formalism.

The scale height of the disk has been explored here for the first time. Its strong impact on the polarization
level together with its extremely light impact on both H$\alpha$ $V/R$ variations and peak separation
might indicate that its influence is limited to the innermost regions of the disk. Based on the polarization
diagnostics, it appears that values of $H_{0}$ higher than those adopted in ZT\,II are favored.

        \subsection{The H$\alpha$ profile: triple-peak or double-dip?}

In the two previous sections, we pointed out the various unusual features observed during the triple-peak phases.
The origin (\emph{\emph{i.e.}}, the physical process giving rise to those profiles) and nature (three distinct emission components
$vs.$ self-absorption overlapping with an emission component) of the triple-peaked profiles are puzzling issues that,
together with the shell-line profile perturbations highlighted here, deserve some discussion. Various systems, like Be/X-ray
binaries \citep{moritani11}, exhibit spectroscopic features that look similar but whose origin might differ from those observed
in $\zeta$\,Tauri. The discussion below likely applies  to objects like $\zeta$\,Tauri, namely Be-shell binaries undergoing
$V/R$ variations, and does not exclude another origin for triple-peaked features observed in other Be stars.

The  spectroscopic diagnostics provide somewhat contradictory clues regarding the origin of the triple-peaked features.
For instance, ZT\,I noted  that other non-hydrogen emission lines in the visual range do not experience a
triple-peak phase. On the one hand, the absence of triple-peaked profiles in optically thin lines like \ion{O}{i}\,8446, which
probes the H$\alpha$-forming region due to Ly$\beta$ resonance pumping \citep{andrillat88}, suggest that its origin is
probably not an actual local density enhancement. Considering the shell nature of $\zeta$\,Tauri, a self-absorption located
at the base of the disk and overlapping with the emission component might be invoked. On the other hand, the absence of
triple-peaked profiles in optically thick lines, like \ion{Fe}{ii}\,6318, suggests that the feature originates from the outer disk,
since those lines are supposed to be formed closer to the star than H$\alpha$ and H$\beta$. This would support a binary
origin, as the companion would more likely affect the outermost regions of the disk.

\begin{table*}[ht]
\centering
\caption{General performance comparison between 2D and 2.5D models.}
\begin{tabular}{c c c c}
\hline
Observable & Probed Region & Related Parameter(s) & Favored Model \\
\hline
\hline
$V/R$ (H$\alpha$)                       & Entire Disk   & $\alpha$, Base Density, Inclination Angle & 2.5D \\
Peak Separation (H$\alpha$)     & Entire Disk   & $\alpha$, Base Density, Inclination Angle & 2D \\
Shell Lines                             & Line of Sight & - & - \\
Polarization                            & Inner Disk    & $\alpha$, Base Density, Inclination Angle, Disk Scale height & 2.5D \\
$V/R$ (\ion{Br}{15})                    & Inner Disk    & $\alpha$ & 2.5D \\
\hline
\end{tabular}
\end{table*}

The results presented in Sects.~3.1 and 3.2 support the idea that, whatever mechanism is creating the triple-peaked profiles,
it should be effective both in the outer disk, from where the emission lines originate, and in the inner disk, from where the shell
lines originate. On that score, the triple-peak phase of Cycle~III is particularly interesting: it is the shortest one,\footnote{It is worth
noticing that Cycle~III is also the shortest $V/R$ cycle, with 1230 days (see Table~2 of ZT\,I).} but also the one where the most
spectacular alterations of the velocity field happen. This would favor a spiral arm-related process rather than a companion-related
one. Indeed, since the extent of the spiral arm gets apparently smaller during Cycle~III (Sect.~3.1), its precession time would
logically follow the same trend, and so would  the duration of its triple-peak phase.

Identifying the nature of the triple-peaked profiles might help us to better understand their origin. Traditionally, disturbed H$\alpha$
profiles like the ones displayed by $\zeta$\,Tauri around $\tau$ = 0.75 have been called triple-peaked profiles, for obvious 
reasons. However, the density modulation induced by a one-armed disk oscillation is very smooth (see Fig.~8). It provides little
inspiration for the delineation of three distinct line-emitting regions. The general success of the model even for $\zeta$\,Tauri
suggests that the presence of the companion does not change this significantly.

However, restricting the search for an interpretation to absorption could lead to an increasing number of additional free parameters
for our models. For instance, there could be self-absorption in a high-density region or the companion star could excite a warping
mode of the disk, which would require not only an extension of our code to 3D, but also an expansion of its dynamical basis. However,  there
is an ansatz, which does not add any new parameters: the double-dip phenomenon,  which it will be discussed here, is most
prominent in H$\alpha$.

Because it is almost a resonance line, H$\alpha$ absorption forms along the line of sight to the central star over a very large range in distance,
larger than for any other Balmer line. If the effect of the one-armed disk  oscillation is pictured as a spiral pattern of density enhancement
(Fig.~8), it is conceivable that the line of sight intersects the increased-density zone of the spiral twice, at about the same azimuth (modulo
360$^{\circ}$) but different radii. Of the resulting two discrete absorption components, one would be roughly shared with the high-order
Balmer lines forming in the inner disk while the latter would exhibit at most traces of the second absorption in H$\alpha$. Because the
spiral is one-armed, double dips only occur once per V/R cycle. This description seems to be in qualitative agreement with the observations of
$\zeta$\,Tau. In this scenario, double dips would  always occur at roughly the same phase in all stars, namely when the outer H$\alpha$-absorbing
end of the spiral crosses the line of sight, which corresponds to the middle of the ascending branch of $V/R$ cycles ($\tau \approx 0.75$).

The apparent restriction of the double-dip phenomenon to shell stars with a companion \citep{rivi06, stefl07} could imply that the companion
not only truncates the disk but also changes its dynamics. For instance, the disk eigenmodes could be affected or the viscosity could be increased,
leading to a more strongly wound spiral pattern (see Fig.~8). Cycle-to-cycle differences in the appearance of double-dip line profiles, as observed in
$\zeta$\,Tau, may supply additional useful interpretative constraints. A forthcoming paper will more specifically investigate the physical basis, if any,
of this picture and explore more quantitatively the diagnostic value of double-dip shell absorptions.

\section{Conclusions}

We undertook the analysis of the Be shell star $\zeta$\,Tauri, introducing new criteria to quantify
the variations of its spectroscopic characteristics. The quantitative analysis of its observational data
was complemented with a detailed modeling, adopting a new formalism based on the results of
\citet{ogilvie08}. The main observational results are the following:

\begin{itemize}
        \item As expected, the $V/R$ variations are accompanied by variations of the peak separation;
        the latter exhibiting a saw-tooth pattern identifiable despite the large scatter among the observational
        points.

        \item Shell lines exhibit marginal variations in depth and more obvious variations in velocity.
        Their apparent synchronism with the $V/R$ variations make them good diagnostic tools for our
        models. Surprisingly, the asymmetry of the shell lines does not vary much. The average value
        for the asymmetry parameter indicates that shell lines are generally broader on their blue side.

        \item Between 1997 and 2010, $\zeta$ Tauri experienced three complete $V/R$ cycles. Each
        cycle exhibits a period during which the emission profiles of H$\alpha$ and H$\beta$ are perturbed,
        making the measurement of the $V/R$ ratio and peak separation uncertain. The origin and nature
        of this so-called triple-peak phase has been discussed and a possible explanation, the double-dip,
        has been suggested. The existence of such features only in Be-shell stars with a companion suggest
        that the physical mechanism creating these profiles may be triggered (or enhanced) by the companion
        itself.

\end{itemize}

From our modeling efforts, we reached the following conclusions (also summarized in Table~3):
        
\begin{itemize}
        \item The new 2.5D global-disk oscillation formalism is generally preferred over the previous 2D model.
        In particular, the 2.5D formalism solves the \ion{Br}{15} phase problem found by \citet{carciofi09} and
        improves the agreement with the observed continuum polarization if a larger disk scale height is
        adopted. This is an encouraging first step toward our forthcoming 3D formalism.

        \item Our models (both 2D and 2.5D) fail to reproduce most of the shell-line properties. They predict
         variations that are too small for the position of the shell absorption component and large asymmetry variations
        that are not observed. They were thus not used as diagnostic tools for constraining model parameters,
        but they highlighted a current limitation of our models that we shall explore in the near future.

        \item As expected by the theory, for a given value of the viscosity parameter $\alpha$, 2.5D models
        produce much more confined spiral density perturbations than their 2D counterparts. Since reproducing
        the $V/R$ variations requires less confined perturbation modes, larger values of $\alpha$ must be adopted
        while computing 2.5D models. This finding is in agreement with the recent determination of a large
        $\alpha$ ($= 1$) for 28\,CMa by \citet{carciofi12}.

        \item The effect on the observables of some model input parameters has been assessed. The base
        density and inclination angle determined in ZT\,II remain valid within the 2.5D formalism. The disk
        vertical scale height has been explored here for the first time; the tests realized on various observables,
        the polarization in particular, tend to favor larger values for that parameter.

\end{itemize}

The present paper is the first  of a series. The second  is already in preparation and aims at applying
the 2.5D model to interferometric data observed with AMBER/VLTI and MIRC/CHARA. As mentioned earlier,
the 2.5D formalism discussed here is a useful step toward the 3D treatment, not the final model. The third
paper of the series will be  dedicated to the 3D modeling of $\zeta$\,Tauri, gathering all the data presented
in the first two papers. The results presented here suggest that the density distribution across the disk is already
correctly reproduced by the 2.5D model, but not the kinematic structure. Therefore, the biggest challenge the
3D model will have to face is reproducing the amplitude of the line-of-sight velocity variations, and providing
 a consistent picture of both the density and kinematic structure of perturbed Be disks.

\begin{acknowledgements}
The authors are very grateful to the anonymous referee for the careful review of the manuscript
 and the useful comments. This work was supported by the FAPESP grant N$^{o}$\,2011/17930-5
 to C.E. and made large use of the computing facilities of the Laboratory of Astroinformatics 
(IAG/USP, NAT/Unicsul), the purchase of which was made possible by the Brazilian agency FAPESP
(grant N$^{o}$\,2009/54006-4) and the INCT-A. C.E. and A.C.C. are grateful to the technical
computing staff of the astronomy department of the IAG/USP for their continuous help. A.C.C.
acknowledges support from CNPq grant N$^{o}$\,307076/2012-1. A.T.O. is grateful to support
from the JSPS grant N$^{o}$\,24540235.
\end{acknowledgements}

\bibliographystyle{aa}
\bibliography{ztau}
\newpage

\appendix

\section{Unperturbed disk state}

Before we consider the equations describing the 3D formulation of \citet{ogilvie08} for the
one-armed oscillation model, we shall note the fundamental equations describing the
unperturbed disk. The model adopted here is the same as that detailed in ZT\,II (their Sect.~2).
We assume a steady, axisymmetric viscous decretion disk, for which the surface density is given by
\citep{bjorkman05}
\begin{equation}
        \Sigma(\varpi)=\frac{\langle \mu \rangle m_{\rm H} {\dot M} {\left( G
        M\right)}^{1/2}} 
        {3 \pi k \alpha \langle T \rangle \varpi^{3/2}}
                \left[(R_0/\varpi)^{1/2}-1\right] \enspace ,
        \label{eq:disk_Sigma}
\end{equation}
where $\varpi=(x^2+y^2)^{1/2}$ is the horizontal distance from the center of the star,
$\langle\mu\rangle$ is the average mean molecular weight, $m_H$ is the mass of the
hydrogen atom, $\alpha$ is the viscosity parameter of \citet{shakura73}, $\langle T\rangle$
is the average gas kinetic temperature, $\dot{M}$ is the stellar decretion rate, and $R_0$
is an arbitrary integration constant.

The mass-decretion rate $\dot{M}$ is related to the radial velocity via the equation
of continuity \citep{carciofi08b}. It is given by
\begin{equation}
        \dot{M} = 2 \pi \varpi \Sigma(\varpi) v_\varpi,
\end{equation}
where $v_\varpi$ is the vertically-averaged, radial velocity.

In binaries with low eccentricities, the Be disk is expected to be truncated by the tidal
torques from the companion \citep{okazaki01b}. For a circular binary with a small mass
ratio  ($q = M_{2}/M_{1} \lesssim 0.3$), such as $\zeta$~Tau, the truncation takes place
at the tidal radius, which is approximately given by $0.9\;R_\mathrm{Roche}$ \citep{whitehurst91},
with $R_\mathrm{Roche}$ being the Roche radius approximately given by \citep{eggleton83}
\begin{equation}
        R_\mathrm{Roche} = a \frac{0.49q^{-2/3}}{0.69q^{-2/3} + \ln (1+q^{-1/3})},
        \label{eq:roche}
\end{equation}
where $a$ is the semi-major axis.

We also assume that the unperturbed disk is in hydrostatic balance in the vertical direction.
Then, the unperturbed density is given by
\begin{eqnarray}
        \rho(\varpi,z)&=&\rho_\mathrm{m}(\varpi) \exp\left(-\frac{z^2}{2H^2}\right)\enspace,
        \label{eq:isothermal_rho}
\end{eqnarray}
where $\rho_\mathrm{m}(\varpi)$ is the disk density at the mid-plane ($z=0$), and the disk 
scale height is given by
\begin{equation}
        H(\varpi) = H_\mathrm{e} \left( \frac{\varpi}{R_\mathrm{e}}\right)^{3/2},
        \label{eq:scaleheight}
\end{equation}
where $H_\mathrm{e} \equiv (k \langle T \rangle R_\mathrm{e})^{1/2}(\langle \mu
\rangle m_\mathrm{H} G M)^{-1/2}$ is the disk scale height at the inner radius 
$\varpi=R_\mathrm{e}$.

We can find the radial dependence of the midplane density by integrating $\rho$ in the $z$ direction:
\begin{equation}
    \Sigma(\varpi) = \int_{-\infty}^\infty \rho(\varpi,z) \, dz 
    = (2\pi)^{1/2}H(\varpi)\rho_\mathrm{m}(\varpi) \enspace .
        \label{eq:sigma_integral}
\end{equation}

From Eqs.~(\ref{eq:disk_Sigma}) to (\ref{eq:sigma_integral}), we have
\begin{equation}
        \rho(\varpi,z) = \frac{\Sigma_\mathrm{e}}{(2\pi)^{1/2}H_\mathrm{e}}
        \left(\frac{R_\mathrm{e}}{\varpi} \right)^{3}
        \left[ \frac{(R_0/\varpi)^{1/2}-1}{(R_0/R_\mathrm{e})^{1/2}-1} \right]
        \exp\left(-\frac{z^2}{2H^2}\right),
\end{equation}
where $\Sigma_\mathrm{e}$ is the surface density at the inner disk radius,
which is written as
\begin{equation}
        \Sigma_0 = \frac{\langle \mu \rangle m_{\rm H} {\dot M} {\left( G
        M\right)}^{1/2}}
        {3 \pi k \alpha \langle T \rangle R_\mathrm{e}^{3/2}}
        \left[(R_0/R_\mathrm{e})^{1/2}-1\right] \enspace .
        \label{eq:disk_Sigma0}
\end{equation}

\section{3D Structure of $m=1$ oscillations in truncated decretion disks}

On the unperturbed state described in Appendix A, we superpose a linear $m = 1$ perturbation
in the form of a normal mode of frequency $\omega$, which varies as $\exp[i(\omega t-\phi)]$.
For simplicity, we assume the perturbation to be isothermal. The linearized perturbed equations
are then obtained as follows:
\begin{equation}
        i(\omega - \Omega) v_{\varpi}^{\prime} + \frac{\partial}{\partial \varpi}\left( v_{\varpi} v_{\varpi}^{\prime} \right)
        - 2\Omega v_{\phi}^{\prime} = - \frac{\partial h^{\prime}}{\partial \varpi},
\end{equation}
\begin{equation}
        i(\omega - \Omega) v_{\phi}^{\prime} + \frac{v_{\varpi}}{\varpi}\frac{\partial}{\partial \varpi}
        \left( \varpi v_{\phi}^{\prime} \right) + \frac{\kappa^{2}}{2\Omega} v_{\varpi}^{\prime}
        = \frac{ih^{\prime}}{\varpi} - \alpha\frac{\partial h^{\prime}}{\partial \varpi},
\end{equation}
\begin{equation}
        i(\omega - \Omega) v_{z}^{\prime} + v_{\varpi}\frac{\partial v_z^{\prime}}{\partial \varpi}
        = - \frac{\partial h^{\prime}}{\partial z},
\end{equation}
\begin{eqnarray}
        i(\omega - \Omega) h^{\prime} &+& v_{\varpi} \frac{\partial h^{\prime}}{\partial \varpi}
        + \frac{h^{\prime}}{\varpi \rho} \frac{\partial}{\partial \varpi} \left( \varpi \rho v_{\varpi} \right)
        + v_{\varpi}^{\prime} \frac{\partial h}{\partial \varpi} + v_{z}^{\prime} \frac{\partial h}{\partial z}  \nonumber \\
        &  & = - c_\mathrm{s}^{2}  \left[ \frac{1}{\varpi} \frac{\partial (\varpi v_{\varpi}^{\prime})}{\partial \varpi}
        - \frac{iv_{\phi}^{\prime}}{\varpi} + \frac{\partial v_{z}^{\prime}}{\partial z} \right].
\end{eqnarray}
Here $c_\mathrm{s}=(k\langle T \rangle/\mu m_\mathrm{H})^{1/2}$ is the isothermal sound speed
and ($v_{\varpi}^{\prime}, v_{\phi}^{\prime}, v_{z}^{\prime}$) the perturbed velocity;
$h=c_\mathrm{s}^2 \ln \rho$ and $h^{\prime}=c_\mathrm{s}^2 \rho'/\rho$ are respectively the
unperturbed and perturbed enthalpy, where $\rho^{\prime}$ is the density perturbation.

Following the formulation made by \citet[][see also \citealt{oktariani09}]{ogilvie08}, in order to take
the 3D effect into account, we must expand perturbed quantities in the $z$-direction
in terms of Hermite polynomials:
\begin{equation}
v_{\varpi}^{\prime} (\varpi,z) = \sum_{n} u_{n}(\varpi)H_{n}(\zeta),
\end{equation}
\begin{equation}
v_{\phi }^{\prime} (\varpi,z) = \sum_{n} v_{n}(\varpi)H_{n}(\zeta),
\end{equation}
\begin{equation}
v_{z}^{\prime} (\varpi,z) = \sum_{n} w_{n}(\varpi)H_{n-1}(\zeta),
\end{equation}
\begin{equation}
h^{\prime} (\varpi,z) = \sum_{n} h_{n}(\varpi)H_{n}(\zeta)
\end{equation}
(\citealt{ogilvie08}; see also \citealt{okazaki87}), where $\mathit{H}_n(\zeta)$ is
the Hermite polynomial defined by
\begin{equation}
H_{n}(\zeta) = \exp \left(\frac{\zeta^{2}}{2} \right) 
   \left(-\frac{d}{d \zeta} \right)^{n} \exp \left(-\frac{\zeta^{2}}{2} \right) ,
\end{equation}
with $\zeta = z/H$ and $n$ being a dimensionless vertical coordinate and a non-negative integer, respectively.
We note that $w_n=0$ for $n=0$.

Replacing the perturbed quantities by their Hermite polynomials expansion, Eqs. (B.1) to (B.4) become
\begin{eqnarray}
        && i(\omega - \Omega)u_n -2\Omega v_n + \frac{\partial}{\partial \varpi} (v_{\varpi} u_n)
        \nonumber \\
        &&\hspace*{2em}   -v_{\varpi}\frac{\partial\ln H}{\partial \varpi} 
        [ n u_n + (n+1)(n+2)u_{n+2} ]
        \nonumber \\
        &&\hspace*{2em}   = -\frac{\partial h_n}{\partial \varpi} + \frac{\partial\ln H}{\partial \varpi}[n h_n +(n+1)(n+2)h_{n+2}],
\label{eq:ogi-48}
\end{eqnarray}
\begin{eqnarray}
        &&i(\omega - \Omega)v_n +\frac{\kappa^2}{2\Omega}u_n \nonumber \\
        &&\hspace*{2em}  +v_{\varpi}\left\{ \frac{1}{\varpi}\frac{\partial}{\partial \varpi}(rv_n) - \frac{\partial \ln H}{\partial \varpi}[nv_n+(n+1)(n+2)v_{n+2}]\right\}
        \nonumber \\
        &&\hspace*{2em}   = \frac{ih_n}{\varpi} \nonumber \\
        &&\hspace*{2em} -\alpha \left\{\frac{\partial h_n}{\partial \varpi} - \frac{\partial \ln H}{\partial \varpi}[ n h_n +(n+1)(n+2)h_{n+2}]\right\},
\label{eq:ogi-49}
\end{eqnarray}
\begin{eqnarray}
        &&i(\omega - \Omega)w_n  \nonumber \\
        &&\hspace*{2em}   +v_{\varpi} \left\{\frac{\partial w_n}{\partial \varpi} - \frac{\partial \ln H}{\partial \varpi} [ n w_n + (n+1)(n+2)w_{n+2}] \right\}
        \nonumber \\
        &&\hspace*{2em}   = -\frac{nh_n}{H},
\label{eq:ogi-50}
\end{eqnarray}
\begin{eqnarray}
        &&i(\omega - \Omega)\frac{h_n}{c_\mathrm{s}^2} + \frac{1}{\varpi\Sigma}\frac{\partial}{\partial \varpi}(\varpi \Sigma u_n) - \frac{iv_n}{\varpi}-\frac{w_n}{H} \nonumber \\
        &&\hspace*{2em} +\frac{v_{\varpi}}{c_\mathrm{s}^2}
        \left[\frac{\partial h_n}{\partial \varpi} + \frac{\partial \ln H}{\partial \varpi} (n h_n + h_{n-2})\right]
        \nonumber \\
        &&\hspace*{2em}   +\frac{\partial \ln H}{\partial \varpi}(nu_n + u_{n-2}) = 0.
\label{eq:ogi-51}
\end{eqnarray}
We note that Eqs.~(\ref{eq:ogi-48})-(\ref{eq:ogi-51}) are reduced to Eqs.~(48)-(51) of \citet{ogilvie08}
provided that the effects of the radial velocity $v_{\varpi}$ and viscosity $\alpha$ are neglected and $i$ is
replaced with $-i$\footnote{This results from the difference between our normal mode form,
$\exp[i(\omega t-\phi)]$, and that of \citet{ogilvie08}, $\exp[i(\phi-\omega t)]$}.

Still following \citet{ogilvie08}, we close the system of resulting equations by assuming $u_{n}\equiv 0$ for
$n\geq 2$. Using the approximations $|\omega| \ll \Omega$, $|\Omega-\kappa| \ll \Omega$,
$(c_{\rm s}/\varpi\Omega)^{2} \ll 1$, and $v_{\varpi} \ll r\Omega$, the basic equations for linear $m=1$
perturbations in viscous decretion disks are then
\begin{eqnarray}
        &&i(\omega - \Omega)u_0 -2\Omega v_0 + \frac{\partial}{\partial \varpi}(v_{\varpi} u_0) \nonumber \\
        &&\hspace*{2em}   = -\frac{\partial h_0}{\partial \varpi} + 2\frac{\partial \ln H}{\partial \varpi}h_2,
\label{eq:u0}
\end{eqnarray}
\begin{eqnarray}
        &&i(\omega - \Omega)v_0 +\frac{\kappa^2}{2\Omega}u_0 + \frac{v_{\varpi}}{\varpi}\frac{\partial}{\partial \varpi}(\varpi v_0)
        \nonumber \\
        &&\hspace*{2em}   = \frac{ih_0}{\varpi} - \alpha\left(\frac{\partial h_0}{\partial \varpi} - 2\frac{\partial \ln H}{\partial \varpi}h_2\right),
\label{eq:v0}
\end{eqnarray}
\begin{eqnarray}
        &&i(\omega - \Omega)\frac{h_0}{c_\mathrm{s}^2} + \frac{1}{\varpi \Sigma}\frac{\partial}{\partial \varpi}(\varpi \Sigma u_0) - \frac{iv_0}{\varpi}
        + \frac{v_{\varpi}}{c_\mathrm{s}^2}\frac{\partial h_0}{\partial \varpi} = 0,
\label{eq:h0}
\end{eqnarray}
\begin{eqnarray}
        &&-i\Omega w_2 = -\frac{2h_2}{H},
\label{eq:w2}
\end{eqnarray}
\begin{eqnarray}
        &&-i\Omega\frac{h_2}{c_\mathrm{s}^2} - \frac{w_2}{H}+\frac{\partial \ln H}{\partial \varpi}u_0 = 0.
\label{eq:h2}
\end{eqnarray}
Eliminating $w_{2}$ and $h_{2}$ from Eqs. (\ref{eq:u0})-(\ref{eq:h2}),  we obtain a set of three
equations for $h_0$, $u_0$, and $v_0$:
\begin{eqnarray}
        &&\frac{\partial h_0}{\partial \varpi} + \left[ i(\omega - \Omega) - 2i\left(\frac{\partial \ln H}{\partial \varpi}\right)^2
        \frac{c_\mathrm{s}^2}{\Omega} \right] u_0 + \frac{\partial}{\partial \varpi} (v_{\varpi} u_0) \nonumber \\
        &&\hspace*{2em} -2\Omega v_0 =0,
\label{eq:basic-1}
\end{eqnarray}
\begin{eqnarray}
        &&\left( -\frac{i}{\varpi}+\alpha \frac{\partial}{\partial \varpi} \right) h_0 + \left[ \frac{\kappa^2}{2\Omega}
        - 2i\alpha \left(\frac{\partial \ln H}{\partial \varpi}\right)^2 \frac{c_\mathrm{s}^2}{\Omega} \right] u_0
        \nonumber \\
        &&\hspace*{2em} + i(\omega - \Omega)v_0 + \frac{v_{\varpi}}{\varpi} \frac{\partial}{\partial \varpi}(\varpi v_0) = 0,
\label{eq:basic-2}
\end{eqnarray}
\begin{eqnarray}
        &&\frac{1}{c_\mathrm{s}^2}\left[i(\omega - \Omega) + v_{\varpi}\frac{\partial}{\partial \varpi}\right] h_0
        + \frac{1}{\varpi \Sigma}\frac{\partial}{\partial \varpi}(\varpi \Sigma u_0)
        \nonumber \\
        &&- \frac{iv_0}{\varpi} = 0.
\label{eq:basic-3}
\end{eqnarray}
Equations (B.19), (B.20), and (B.21) correspond respectively to Eqs. (16), (17), and (15) from ZT\,II.
They are solved with the same boundary conditions, \emph{\emph{i.e.}}, a rigid wall $(u_0, v_0) = 0$ at the inner
disk radius and $\Delta\,p = 0$ at the outer disk radius, $\Delta\,p$ being the Lagrangian perturbation of
pressure.

We note that the terms $-2i(\partial \ln H/\partial \varpi)^2 c_\mathrm{s}^2/\Omega$ ($= -9ic_\mathrm{s}^{2}/2r^{2}\Omega$
in isothermal disks) in Eq.~(\ref{eq:basic-1}) and $-2i\alpha (\partial \ln H/\partial \varpi)^2 c_\mathrm{s}^2/\Omega$
in Eq.~(\ref{eq:basic-2}) result from the 3D effect and provides an important contribution to
the confinement of the $m = 1$ modes. The other terms are the same as in the basic equations used in ZT\,II.

Once  Eqs.~(\ref{eq:basic-1})-(\ref{eq:basic-3}) are solved, the second-order terms, $w_2$ and $h_2$,
are obtained from Eqs.~(\ref{eq:w2}) and (\ref{eq:h2}),
\begin{equation}
        w_2 = 2\frac{\partial \ln H}{\partial \varpi}\frac{c_\mathrm{s}}{\Omega}u_0,
\label{eq:w2f}
\end{equation}
\begin{equation}
        h_2 = \frac{\partial \ln H}{\partial \varpi}\frac{c_\mathrm{s}^2}{\Omega}iu_0,
\label{eq:h2f}
\end{equation}
where $c_\mathrm{s} = \Omega H$.
Finally, under the closure approximation adopted here, the perturbed density and velocity are given by
\begin{eqnarray}
        \rho^{\prime}(\varpi, \phi, z) &=& \Re \left\{\rho(\varpi, z) \left[ \frac{h_0}{c_\mathrm{s}^2}H_0\left(\frac{z}{H}\right)
        + \frac{h_2}{c_\mathrm{s}^2}H_2\left(\frac{z}{H}\right)\right]e^{i(\omega t-\phi)} \right\} \nonumber \\
        &=& \rho(\varpi, z) \nonumber \\
        &\times& \left\{ \left\{\Re\left(\frac{\rho_0}{\rho}\right) + \Re\left(\frac{\rho_2}{\rho}\right)
        \left[\left(\frac{z}{H}\right)^2-1\right] \right\}
        \cos (\omega t - \phi ) \right. \nonumber \\
        &-& \left\{ \Im \left(\frac{\rho_0}{\rho}\right) + \Im\left(\frac{\rho_2}{\rho}\right)
        \left[\left(\frac{z}{H}\right)^2-1\right] \right\} \nonumber \\
        && \times \sin (\omega t - \phi ) \left. \right\}, 
\label{eq:rhoprime} \\
        v_{\varpi}^{\prime}(\varpi, \phi, z) &=& \Re \left\{u_0 H_0\left(\frac{z}{H}\right) e^{i(\omega t-\phi)} \right\} \nonumber \\
        &=& \Re \left(u_0\right) \cos (\omega t - \phi ) - \Im \left(u_0\right) \sin (\omega t - \phi ),
\label{eq:uprime} \\
        v_{\phi}^{\prime}(\varpi, \phi, z)  &=& \Re \left\{v_0 H_0\left(\frac{z}{H}\right) e^{i(\omega t-\phi)} \right\} \nonumber \\
        &=& \Re\left(v_0\right) \cos (\omega t - \phi ) - \Im\left(v_0\right) \sin (\omega t - \phi ),
\label{eq:vprime} \\
        v_{z}^{\prime}(\varpi, \phi, z)  &=& \Re \left\{w_2 H_1\left(\frac{z}{H}\right) e^{i(\omega t-\phi)} \right\} \nonumber \\
        &=& \left[ \Re\left(w_2\right) \cos (\omega t - \phi ) - \Im\left(w_2\right) \sin (\omega t - \phi )  \right] \nonumber \\
        &\times& \frac{z}{H},
\label{eq:wprime}
\end{eqnarray}
where the unperturbed density $\rho(\varpi, z)$ is given by Eq.~(\ref{eq:isothermal_rho}) and the density perturbation 
is related to the enthalpy perturbation by $\rho_0/\rho=h_0/c_\mathrm{s}^2$ and $\rho_2/\rho=h_2/c_\mathrm{s}^2$.

Equations (B.24) to (B.27) provide the 3D density and kinematic structure of the perturbed disk. In  the framework of
the 2.5D formalism, the second-order terms, namely $w_{2}$ and $\rho_{2}$, are taken as equal to zero. Consequently, the
$z$ component of the velocity field ($v_{z}^{\prime}$) and the $z$-dependent terms of the perturbed density ($\rho^{\prime}$)
are neglected. The final solution is thus independent of the vertical coordinate and includes zeroth-order terms only.

\end{document}